\documentclass[a4paper,fleqn]{cas-dc}

\usepackage[authoryear]{natbib}

\usepackage{amsmath,amssymb,amsfonts}
\usepackage{algorithmic}
\usepackage{graphicx}
\usepackage{textcomp}
\usepackage{xcolor}
\usepackage{ragged2e}
\usepackage{balance}
\usepackage{colortbl}
\usepackage{multirow}
\usepackage{balance}
\usepackage[figuresright]{rotating}
\usepackage{rotfloat}
\usepackage{color}

\usepackage{CJKutf8}

\usepackage{hyperref}

\hypersetup{
    colorlinks=true,
    linkcolor=blue,
    anchorcolor=blue,
    citecolor=blue
}

\def\tsc#1{\csdef{#1}{\textsc{\lowercase{#1}}\xspace}}
\tsc{WGM}
\tsc{QE}
\tsc{EP}
\tsc{PMS}
\tsc{BEC}
\tsc{DE}
\begin{document}
\begin{sloppypar}

\let\WriteBookmarks\relax
\def\floatpagepagefraction{1}
\def\textpagefraction{.001}

\shorttitle{Exploring Multi-Programming-Language Commits and Their Impacts on Software Quality}
\shortauthors{Z Li et al.}
%\title [mode = title]{Multi-Programming-Language Commits in Open Source Software: An Empirical Study on Apache Projects}
\title [mode = title]{Exploring Multi-Programming-Language Commits and Their Impacts on Software Quality: An Empirical Study on Apache Projects}
%\tnotetext[1]{This work is supported by the Natural Science Foundation of Hubei Province of China under Grant No. 2021CFB577, and the National Natural Science Foundation of China under Grant Nos. 62176099 and 62002129.}

\author[1]{Zengyang Li}
\ead{zengyangli@ccnu.edu.cn}

\credit{Conceptualization of this study, Methodology, Investigation, Data curation, Writing - Original draft preparation}

\address[1]{School of Computer Science \& Hubei Provincial Key Laboratory of Artificial Intelligence and Smart Learning, \\Central China Normal University, Wuhan, China \\}

\author[1]{Xiaoxiao Qi}
\ead{qixiaoxiao@mails.ccnu.edu.cn}
\credit{Investigation, Data curation, Software, Writing - Original draft preparation}

\author[1]{Qinyi Yu}
\ead{qinyiyu@mails.ccnu.edu.cn}
\credit{Investigation, Data curation, Software}

\author[2]{Peng Liang}
\cormark[1]
\ead{liangp@whu.edu.cn}
\credit{Conceptualization of this study, Methodology, Writing - Original draft preparation}
\address[2]{School of Computer Science, Wuhan University, Wuhan, China}

\author[1]{Ran Mo}
\ead{moran@ccnu.edu.cn}
\credit{Methodology, Writing - Original draft preparation}

\author[3]{Chen Yang}
\ead{yangchen@szpt.edu.cn}
\credit{Methodology, Writing - Original draft preparation}
\address[3]{School of Artificial Intelligence, Shenzhen Polytechnic, Shenzhen, China}

\cortext[cor1]{Corresponding author.}

\begin{abstract}
\noindent \textbf{Context}: Modern software systems (e.g., Apache Spark) are usually written in multiple programming languages (PLs). There is little understanding on the phenomenon of multi-programming-language commits (MPLCs), which involve modified source files written in multiple PLs.\\
\textbf{Objective}: This work aims to explore MPLCs and their impacts on development difficulty and software quality.\\
\textbf{Methods}: We performed an empirical study on eighteen non-trivial Apache projects with 197,566 commits.\\
\textbf{Results}: (1) the most commonly used PL combination consists of all the four PLs, i.e., C/C++, Java, JavaScript, and Python; (2) 9\% of the commits from all the projects are MPLCs, and the proportion of MPLCs in 83\% of the projects goes to a relatively stable level; (3) more than 90\% of the MPLCs from all the projects involve source files in two PLs; (4) the change complexity of MPLCs is significantly higher than that of non-MPLCs; (5) issues fixed in MPLCs take significantly longer to be resolved than issues fixed in non-MPLCs in 89\% of the projects; (6) MPLCs do not show significant effects on issue reopen; (7) source files undergoing MPLCs tend to be more bug-prone; and (8) MPLCs introduce more bugs than non-MPLCs.\\
\textbf{Conclusions}: MPLCs are related to increased development difficulty and decreased software quality.

\end{abstract}

%\includegraphics{figs/grabs.pdf}

% \item The first attempt to comprehensively explore the phenomenon of Multi-Programming-Language Commits (MPLCs).
% \item The change complexity of MPLCs is significantly higher than that of non-MPLCs.
% \item Issues fixed in MPLCs take significantly longer to be resolved than issues fixed in non-MPLCs in 80\% of the projects.
% \item Source files that have been modified in MPLCs tend to be more bug-prone than source files that have never been modified in MPLCs.
% \item MPLCs are more likely to introduce bugs than non-MPLCs, and a bug-introducing MPLC introduces more bugs than a bug-introducing non-MPLC.


\maketitle

\section{Introduction}
\label{chap:intro}
Modern software systems, such as Apache Ambari and Spark, are usually written in multiple programming languages (PLs). One main reason for adopting multiple PLs is to reuse existing code with required functionalities \citep{GrAbJa2020}. Another main reason is to take advantages of specific PLs to implement certain features, to meet various software quality needs, and to improve software development efficiency \citep{GrAbJa2020, RaPoFiDe2014, MaBa2015,KoWiLo2016,Ma2017,MaKiLe2017,AbGrKh2019}. Nowadays, multi-programming-language (MPL) software development is increasingly prevalent with the technology advances \citep{KoLiWo2006,KoWiLo2016,AbRaOpKh2021}. 

However, despite great benefits of MPL systems, they are also facing various challenges. For instance, static code analysis is much more difficult in MPL systems than mono-language systems since multiple PLs and cross-language communication mechanisms need to be analyzed simultaneously \citep{ShMiAb2019, kaIsIz2019}.
%\red{For instance, the Apache Spark project is written in multiple PLs (including Scala, Java, C/C++, Python, JavaScript, and Ruby), and it also provides multi-programming-language (MPL) interfaces of Scala, Java, and Python to fulfill the requirement of big data processing.}
%\red{\begin{CJK}{UTF8}{gbsn}[存在问题和挑战，研究者试图从XX方面去解决，但有XX问题还没有解决，或需要进一步研究.]\end{CJK}}.

In an MPL system, there inevitably exist a certain proportion of code changes that need to modify source files written in different PLs. Intuitively, a code change in which the modified source files are written in multiple PLs is likely to modify more than one component in a software system, and thus the complexity of such a change is relatively high. As a result, such a code change may need more effort and time to understand and analyze the impacts on the parts affected by the modified source files. We define an MPL commit (MPLC) in a version control system (e.g., Git) as a commit that involves modified source files written in two or more PLs. 

Although there are a few studies that investigated the quality of MPL systems \citep{RaPoFiDe2014, KoWiLo2016, GrEgAd2020}, these studies took a project or pull request as an analysis unit, which is at a relatively high level and may not provide specific advice for software development practice. In contrast, a commit is in a finer granularity than a project or pull request, and developers deal with code changes in commits in daily development practices. Hence, we suggest to investigate MPL software development from the perspective of commits (i.e., MPLCs). To the best of our knowledge, the phenomenon of MPLCs in software development has not been comprehensively explored yet. 
In light of the potential impacts of MPLCs on development difficulty (e.g., cross-language change impact analysis \citep{ShMiAb2019}) and software quality (e.g., bug introduction \citep{KoWiLo2016, BeHoMaViVi2019}), there is a need for exploring the phenomenon of MPLCs. To this end, we decided to conduct an empirical study in order to understand the state of MPLCs, their change complexity, as well as their impacts on open time of issues, issue reopen, bug proneness of source files, and bug introduction in real-life software projects, which provides a foundation for improving the practices of MPL software development.

The main contributions of this paper are summarized as follows:
 
 \begin{itemize}
 \setlength{\itemsep}{0pt}
 \setlength{\parsep}{0pt}
  \item This work is a first attempt to comprehensively explore the phenomenon of MPLCs in real-world settings.
  \item The state of MPLCs (including the proportion of MPLCs and the number of PLs used in MPLCs) in MPL software systems is explored. 
  \item In this work, we investigated in depth the change complexity of MPLCs, open time of issues fixed in MPLCs, bug proneness of source files modified in MPLCs, as well as the impacts of MPLCs on issue reopen and bug introduction in MPL software systems.
\end{itemize}

This paper is an extended version of our previous conference paper \citep{LiQiYu2021} published in the \emph{Proceedings of the 29th IEEE/ACM International Conference on Program Comprehension (ICPC 2021)}. The main extension of this journal version compared to the previous conference version lies in the following aspects: (1) we added three research questions on the use of PLs and the impacts of MPLCs on issue reopen and bug introduction; (2) we enriched the results of the research question on the proportion of MPLCs by calculating the correlation between the proportion of MPLCs and a measure (i.e., the entropy of PL use) reflecting how balanced PLs are used in the selected projects; and (3) we discussed the implications of results and threats to validity more comprehensively.

The remaining of this paper is organized as follows. Section \ref{chap:relat} presents the related work; Section \ref{chap:case} describes the design of the empirical study; Section \ref{chap:study} presents the results of the study; Section \ref{chap:discus} discusses the study results; Section \ref{chap:threats} identifies the threats to validity of the results; and Section \ref{conclusions} concludes this work with future research directions.

\section{Related Work}\label{RelatedWork}
\label{chap:relat}

{To the best of our knowledge, there has not been work on MPLCs. Thus, the related work presented here is not directly relevant to MPLCs, but related to the research on MPL software systems in general. The related work is presented in two aspects, including the phenomenon of MPL software systems and quality of MPL software systems.}

\subsection{Phenomenon of MPL Software Systems}\label{RelatedWork_A}
Mayer and Bauer studied the phenomenon of multi-language programming using data mining technologies on 1,150 open source software (OSS) projects gathered from GitHub \citep{MaBa2015}. They used the Poisson regression model to explore the relationship between the number of PLs of the project and the size, age, number of contributors, and number of commits. They found that each project uses an average of 5 PLs with a clearly dominant PL; the median number of general-purpose PLs and domain-specific PLs is 2 and 2, respectively. The results also confirmed that the use of multiple PLs is very common in OSS projects. The focus of our work is different in that we investigated MPL software systems from the perspective of commits, while the work of Mayer and Bauer paid more attention at the level of project; in addition, we went deeper and looked into the bug proneness and change complexity of source files modified in MPLCs as well.

\subsection{Quality of MPL Software Systems}\label{RelatedWork_B}
In 2011, Bhattacharya and Neamtiu investigated the impact of C and C++ on software maintainability and code quality of four large OSS projects \citep{BhNe2011}. They found that C++ code has higher internal quality than C code and C++ code is less prone to bugs than C code, but they could not confirm that C++ code needs less effort to maintain than C code. Their work looked into the impact of specific PLs on code quality, while our work investigated the impact of MPLCs on the bug proneness of source files in terms of defect density.

In 2014, Ray \emph{et al}. studied the effect of PLs on software quality using a large dataset of 729 projects in 17 PLs gathered from GitHub \citep{RaPoFiDe2014}. They combined multiple regression modeling with visualization and text analytics to study the impact of language characteristics. They found that language design indeed has a significant but modest effect on software defects. In addition, they found that there is a small but significant correlation between language set and software defects. Specifically, they found that there are 11 PLs that have a relationship with software defects. In 2019, Berger \emph{et al}. \citep{BeHoMaViVi2019} carried out repeated experiments of the study of Ray \emph{et al}. \citep{RaPoFiDe2014} and reduced the number of defect-related languages down from 11 to only 4. These studies paid attention to the impact of language features on bug proneness of software systems. In contrast, our work is focused on the impact of MPLCs on the bug proneness of source files.

In 2016, Kochhar \emph{et al}. conducted a large-scale empirical investigation of the use of multiple PLs and the combination of certain PLs on bug proneness \citep{KoWiLo2016}. They analyzed a dataset comprised of 628 projects collected from GitHub, in 17 general-purpose PLs (e.g., Java and Python). They found that implementing a project with more PLs significantly increases bug proneness, especially on memory, concurrency, and algorithm bugs. The results also revealed that the use of specific PLs together is more bug-prone in an MPL setting. However, our work is focused on the development difficulty and software quality in a finer granularity of commits.

In 2019, Abidi \emph{et al}. identified six anti-patterns \citep{AbKhGu2019a} and twelve code smells \citep{AbKhGu2019b} in MPL software systems. Six anti-patterns were identified in OSS systems, including excessive inter-language communication, too much scattering, language and paradigms mismatch, and so forth \citep{AbKhGu2019a}. Twelve code smells were proposed, including passing excessive objects, not handling exceptions across languages, assuming safe multi-language return values, memory management mismatch, and so on \citep{AbKhGu2019b}. Abidi \emph{et al}. subsequently proposed an approach to detect aforementioned anti-patterns and code smells (both called design smells according to the authors) in MPL systems in which Java Native Interface (JNI) is used, and conducted an empirical study on the fault proneness of such MPL design smells in nine open source JNI projects \citep{AbRaOpKh2021}. They found that MPL design smells are prevalent in the selected projects and files with MPL design smells can often be more associated with bugs than files without these design smells, and that specific smells are more correlated to fault proneness than other smells. These design smells provide useful suggestions in practice to avoid design defects and implementation flaws in software development. 

In 2019, Kargar \emph{et al}. proposed an approach to modularization of MPL applications \citep{kaIsIz2019}. The results show that the proposed approach can build a modularization close to human experts, which may be helpful in understanding MPL software systems. In 2020, the same authors proposed a method to improve the modularization quality of heterogeneous MPL software systems by unifying structural and semantic concepts \citep{kaIsIz2020}. Admittedly, architecture quality (e.g., modularity) of MPL software systems is worth further and deeper research. This study provided an important viewpoint for the research on MPL software systems. Based on the results of our work, we will examine the architecture quality of MPL software systems with a relatively high proportion of MPLCs.

In 2020, Grichi \emph{et al}. performed a case study on the impact of interlanguage dependencies in MPL systems \citep{GrAbJa2020}. They found that the risk of bug introduction gets higher when there are more interlanguage dependencies, while this risk remains constant for intralanguage dependencies; the percentage of bugs found in interlanguage dependencies is three times larger than the percentage of bugs identified in intralanguage dependencies.
Grichi \emph{et al}. also conducted a study on the impact of MPL development in machine learning frameworks \citep{GrEgAd2020}. They found that mono-language pull requests in machine learning frameworks are more bug-prone than traditional software systems. Their work investigated the influence of MPL code changes on software systems in a granularity of pull requests, while our work studied the bug proneness of MPL code changes on software systems in a finer granularity of commits. In addition, our work demonstrates a higher bug proneness of source files modified in MPLCs than source files modified in only non-MPLCs, which is a dramatic difference from the results obtained by the work of Grichi \emph{et al}. 

In 2021, Shen \emph{et al}. carried out a preliminary investigation regarding cross-language coupling detection on Android applications. They found that the code changes in commits involving Jave/Kotlin and XML files are more scattered and more likely to introduce bugs than other commits. This finding is in line with our result. However, only the commits containing both Jave/Kotlin and XML were considered in their work, while MPLCs involving no less than two of eighteen general-purpose PLs are studied in our work.

\section{Case Study Design}
\label{chap:case}
In order to investigate the state of MPLCs and their impacts on development difficulty and software quality, we performed a case study on Apache OSS projects. The main reason for conducting a case study is that, through using OSS projects, and more specifically their commit records and issues, we examine the phenomenon in its real-life context, since both the commit records and issues cannot be monitored in isolation, and their environment cannot be controlled. In this section we describe the case study, which was designed and reported following the guidelines proposed by Runeson and H{\"o}st \citep{RuHo2009}.
\subsection{Objective and Research Questions}\label{DesignRQ}
The goal of this study, described using the Goal-Question-Metric (GQM) approach \citep{Ba1992}, is: to \emph{analyze} commits as well as their corresponding modified source files and fixed issues \emph{for the purpose of} exploration \emph{with respect to} the state of MPLCs as well as their impacts on development difficulty and software quality, \emph{from the point of view of} software developers \emph{in the context of} MPL OSS development.

Based on the abovementioned goal, we have formulated eight research questions (RQs), which are classified into three categories and described as follows. 

~\\
\noindent \textbf{Category I}: State of MPLCs.

\begin{itemize}
\setlength{\itemsep}{0pt}
\setlength{\parsep}{0pt}
\item [\textbf{RQ1:}] How do the programming languages distribute over the selected projects? \\
\textbf{Rationale}: With this RQ, we investigate what PLs are used and how balanced the PLs are used in the MPL projects. In addition, we can also find out the most commonly-used combination of PLs.
\item[\textbf{RQ2:}] What is the proportion of MPLCs over all commits of a project? Is the proportion of MPLCs correlated with the degree of the balanced use of PLs in all the projects?\\
\textbf{Rationale:} With this RQ, we investigate the frequency of MPLCs occurred in software projects and how the proportion of MPLCs evolves, so as to get a basic understanding on the state of MPLCs in MPL software projects. In addition, we study the relationship between the proportion of MPLCs and the distribution of PLs used in the selected projects, in order to get a further understanding on the potential influence of the use of PLs on the MPLCs of a project.
\item[\textbf{RQ3:}] How many programming languages are used in the modified source files of MPLCs?\\
\textbf{Rationale:} This RQ is focused on the number of PLs used in source files modified in MPLCs, which enables us to understand the tendency of the use of multiple PLs.
\end{itemize}

\noindent \textbf{Category II}: Impacts of MPLCs on development difficulty.

\begin{itemize}
\setlength{\itemsep}{0pt}
\setlength{\parsep}{0pt}

\item[\textbf{RQ4:}] What is the code change complexity of MPLCs? Is there a difference on the code change complexity between MPLCs and non-MPLCs?\\
\noindent \textbf{Rationale}: To explore the development difficulty of MPLCs, we look into the code change complexity of MPLCs. Intuitively, the complexity of code changes in MPLCs may be different from that in non-MPLCs. With this RQ, we intend to calculate the complexity of code changes in MPLCs and further validate if the complexity of code changes in MPLCs is significantly higher than that in non-MPLCs. In addition, the complexity of code change in a commit can be measured by the number of lines of code, source files, and directories that are modified in the commit, and by the entropy \citep{Ha2009} of the modified files in the commit. These change complexity measures are adopted from \citep{LiLiLiMoLi2020}.

\item[\textbf{RQ5:}] Do the issues fixed in MPLCs tend to take longer to be resolved than issues fixed in non-MPLCs?\\
\textbf{Rationale}: With this RQ, we further investigate the time taken to resolve issues that were fixed in MPLCs and non-MPLCs. The time taken to resolve an issue can, to some extent, reflect the development difficulty of MPL software systems.

\end{itemize}

\noindent \textbf{Category III}: Impacts of MPLCs on software quality.

\begin{itemize}
\setlength{\itemsep}{0pt}
\setlength{\parsep}{0pt}

\item[\textbf{RQ6:}] Are issues fixed in MPLCs more likely to be reopened than issues fixed in non-MPLCs?\\
\textbf{Rationale}: With this RQ, we investigate the likelihood of reopening issues fixed in MPLCs, which reflects the analyzability of the source code of files involved in MPLCs.

\item[\textbf{RQ7:}] Are source files that have been modified in MPLCs more bug-prone than source files that have never been modified in MPLCs?\\
\textbf{Rationale}: MPLCs may influence the quality of software systems. With this RQ, we are concerned with the impact of MPLCs on software systems in terms of the likelihood of bugs.

\item[\textbf{RQ8:}] Are MPLCs more likely to introduce bugs than non-MPLCs?\\
\textbf{Rationale}: With this RQ, we look into the impact of MPLCs on bug introduction at the granularity of commit. In contrast, with RQ7 we study the influence of MPLCs on bug proneness at the granularity of source file.
\end{itemize}

Among the eight RQs, RQ1, the second part of RQ2, RQ6, and RQ8 are new RQs addressed in this journal version, comparing with our previous conference paper \citep{LiQiYu2021}.

\subsection{Cases and Unit Analysis }\label{CasesandUnitAnalysis}
According to \citep{RuHo2009}, case studies can be characterized based on the way they define their cases and units of analysis. This study investigates multiple MPL OSS projects, i.e., cases, and each commit and the corresponding issue fixed is a single unit of analysis.

\begin{table}[]
\caption{Programming languages examined.}
\centering
\begin{tabular}{|c|c|c|c|c|c|}
\hline
\textbf{\#} & \textbf{PL}  & \textbf{\#} & \textbf{PL} & \textbf{\#} & \textbf{PL} \\ \hline
1           & C/C++        & 7           & Haskell     & 13          & PHP         \\ \hline
2           & C\#          & 8           & Java        & 14          & Python      \\ \hline
3           & Clojure      & 9           & JavaScript  & 15          & Ruby        \\ \hline
4           & CoffeeScript & 10          & Kotlin      & 16          & Scala       \\ \hline
5           & Erlang       & 11          & Objective-C & 17          & Swift  \\ \hline
6           & Go           & 12          & Perl        & 18          & TypeScript       \\ \hline
\end{tabular}
\label{table:ProgrammingLanguages}
\end{table}

\subsection{Case Selection}\label{CaseSelection}
In this study, we only investigated Apache MPL OSS projects. The reason why we used Apache projects is that the links between issues and corresponding commits tend to be well recorded in the commit messages of those projects. For selecting each case (i.e., MPL OSS project) included in our study, we applied the following inclusion criteria:

\begin{itemize}
\setlength{\itemsep}{0pt}
\setlength{\parsep}{0pt}
 \item[\textbf{C1:}] The issues of the project are tracked in JIRA. This criterion was set to ensure the same format of all issues of different projects so that we can handle the issues in the same way.
  \item[\textbf{C2:}] No less than 3 out of the 18 PLs listed in Table \ref{table:ProgrammingLanguages} are used in the project. All the 18 listed PLs are general-purpose languages. Sixteen out of the 18 PLs were adopted from \citep{KoWiLo2016}, in which C and C++ are different PLs. However, we combined C and C++ into a single PL in this work since we cannot determine a header file with the extension of “.h” as a source file of C or C++ by only checking the commit records. Besides, we added two general-purpose PLs, i.e., Kotlin and Swift. This criterion was set to enlarge the population of MPLCs with different numbers of PLs and to reduce the dominance of MPLCs of two PLs, thereby increasing the generalizability of the study results.
  \item[\textbf{C3:}] The source code written by the main PL is no more than 75\% of the code of the project. This criterion was set to ensure that the PL use is not extremely unbalanced so that the biases caused by specific PLs can be reduced.
  \item[\textbf{C4:}] The project has more than 20 contributors. This criterion was set to ensure that the selected projects are written by a relatively wide range of developers, thereby reducing biases caused by a small number of developers to the study results.
  \item[\textbf{C5:}] The project has more than 2,000 commits. This criterion was set to ensure that the selected project is non-trivial, and the resulting dataset is big enough to be statistically analyzed.
  \item[\textbf{C6:}] The number of issues of the project is no less than 1,000. This criterion was set to ensure that the selected projects had been in the maintenance and evolution stage for a reasonable length of period, and thus sufficient data on issue open time, issue reopen, bug proneness, and bug introduction can be collected.
  \item[\textbf{C7:}] The project has more than 100 MPLCs. This criterion was set to ensure that the resulting dataset is big enough to be statistically analyzed.
\end{itemize}

\begin{table*}[]
\caption{Data items to be collected for each commit.}
 \centering
\begin{tabular}{|l|l|l|l|}
\hline
\textbf{\#} & \textbf{Name} & \textbf{Description}                                                                                                               & \textbf{RQ}                                        \\ \hline
D1          & CmtID         & The hashcode of the commit.                                                                                                            & -                                                  \\ \hline
D2          & CmtDT         & \begin{tabular}[c]{@{}l@{}}The date and time when the commit happened.\end{tabular}                                                 & -                                                  \\ \hline
D3          & Committer     & The committer of the commit.                                                                                                       & -                                                  \\ \hline
D4          & IsMPLC        & Whether the commit is an MPLC.                                                          & RQ1-RQ8 \\ \hline
D5          & PLNo          & The number of PLs used in the modified source files of the commit. & RQ3                                                \\ \hline
D6          & LOCM          & The number of lines of source code modified in the commit.                             & RQ4                                                \\ \hline
D7          & NOFM          & \begin{tabular}[c]{@{}l@{}}The number of source files modified in the commit.\end{tabular}                                      & RQ4                                                \\ \hline
D8          & NODM          & \begin{tabular}[c]{@{}l@{}}The number of directories modified in the commit.\end{tabular}                                       & RQ4                                                \\ \hline
D9         & Entropy      & \begin{tabular}[c]{@{}l@{}}The normalized entropy of the modified files in the commit \citep{Ha2009}.\end{tabular}                                       & RQ4                                                \\ \hline
D10         & IssueID       & \begin{tabular}[c]{@{}l@{}}The ID of the issue fixed in the commit if applicable.\end{tabular}                                  & RQ5, RQ7 \\ \hline
D11         & IssueRT       & The reporting time of the issue.                                                                                                   & RQ5                                                \\ \hline
D12         & IssueCT       & \begin{tabular}[c]{@{}l@{}}The closing or resolving time of the issue.\end{tabular}                                             & RQ5                                                \\ \hline
D13         & IssueType     & The type (e.g., bug) of the issue.                                                                                                 & RQ7                                                \\ \hline
D14         & IntroducedBugs     & The bugs introduced by the commit.                                                                                                 & RQ8                                                \\ \hline
D15         & IsIssueReopened     & Whether the issue was reopened.                                                                                                 & RQ6                                                \\ \hline
\end{tabular}
\label{table:DataItemsForCommit}
\end{table*}

\begin{table}[]
\caption{Data items to be collected for each source file.}
\centering
\begin{tabular}{|p{0.07\columnwidth}|p{0.1\columnwidth}|p{0.52\columnwidth}|p{0.07\columnwidth}|}
\hline
\textbf{\#} & \textbf{Name} & \textbf{Description}                                                                            & \textbf{RQ} \\ \hline
D16       & Path          & The path of the source file.                                                                    & RQ7         \\ \hline
D17       & LOC           & The number of lines of code in the source file.      & RQ7         \\ \hline
D18       & NOB           & The number of bugs that the source file experienced. & RQ7         \\ \hline
D19       & PLF           & The PL used in the source file. & RQ1         \\ \hline
\end{tabular}
\label{table:DataItemsForFile}
\end{table}

\subsection{Data Collection}\label{DataCollection}
\subsubsection{Data Items to be Collected}
\label{dataitems}
To answer the RQs, we took a commit as the unit of analysis and the data items to be collected are listed in Table \ref{table:DataItemsForCommit}; and to answer RQ7, we also needed to collect the data items described in Table \ref{table:DataItemsForFile} of each source file, which can be extracted from the commits containing the source file. All the data items to be collected except for D9 and D14 are straightforward, thus we only explain these two data items in detail. 

First, we explain the definition of the entropy of the modified source files in a commit (i.e., D9) \citep{Ha2009}. Suppose that the modified source files of commit $c$ are  $\{f_1,f_2,\cdots,f_n\}$, and file $f_i\left(1\leq i\leq n\right)$ was modified in $m_i$ commits during a period of time  before the commit. Let 
\begin{equation}
p_i = m_i/\sum_{i=1}^nm_i .
\end{equation}
\noindent Then, the entropy 
\begin{equation}
H(n) = -\sum_{i=1}^np_ilog_2 p_i .
\end{equation}
\noindent Since the number of modified source files differs between different periods, we need to normalize the entropy to be comparable. Considering that $H(n)$ achieves the maximum of $log_2 n$ when $p_i=1/n$ $(1\leq i\leq n)$, the normalized entropy
\begin{equation}
\widetilde{H}(n)=
\begin{cases}
H(n)/log_2 n & n>1 ,\\
0 &n=1 .
\end{cases}
\end{equation}
In this study, the period is set to 60 days (including the day when commit $c$ happened), which is chosen according to \citep{LiLiLiMoLi2020}.

In the next, we explain what can be considered as introduced bugs by a commit (i.e., D14). In this study, we adopted the SZZ algorithm \citep{SZZ2005} to identify the bug-introducing commits of each bug that is explicitly mentioned in a specific commit message, and then we can collect the bugs introduced by each commit. The SZZ algorithm is widely used in software engineering and research fields, such as defect prediction techniques and bug localization. PyDriller\footnote{https://github.com/ishepard/pydriller} provides a set of APIs for the SZZ algorithm, which is based on AG-SZZ (a variant of SZZ) \citep{Spadini2018}. It can be easily used to identify bug-introducing commits by using the API \textit{get\_commits\_last\_modified\_lines}. In this way, we can get a mapping table between commits and their introduced bugs. Thus, the introduced bugs of each commit can be obtained.
%The steps are as follows: (i) filter out the commit which are not associated with a issue or the type of issue it fixed is not “bug”. (ii) use PyDriller to identify bug-introducing commit. (iii) mark the table according to the mapping between commits and bugs. Similarly, From the bug perspective, we marked the commit(s) that caused the bug (BIC) and whether the bug was caused by MPLC (isMPLC). Specifically, if one of BIC is MPLC, we considered the bug is caused by MPLC.

\subsubsection{Data Collection Procedure}\label{datacollectionprocedure}
\begin{figure}
\centerline{\includegraphics[width=3.0in]{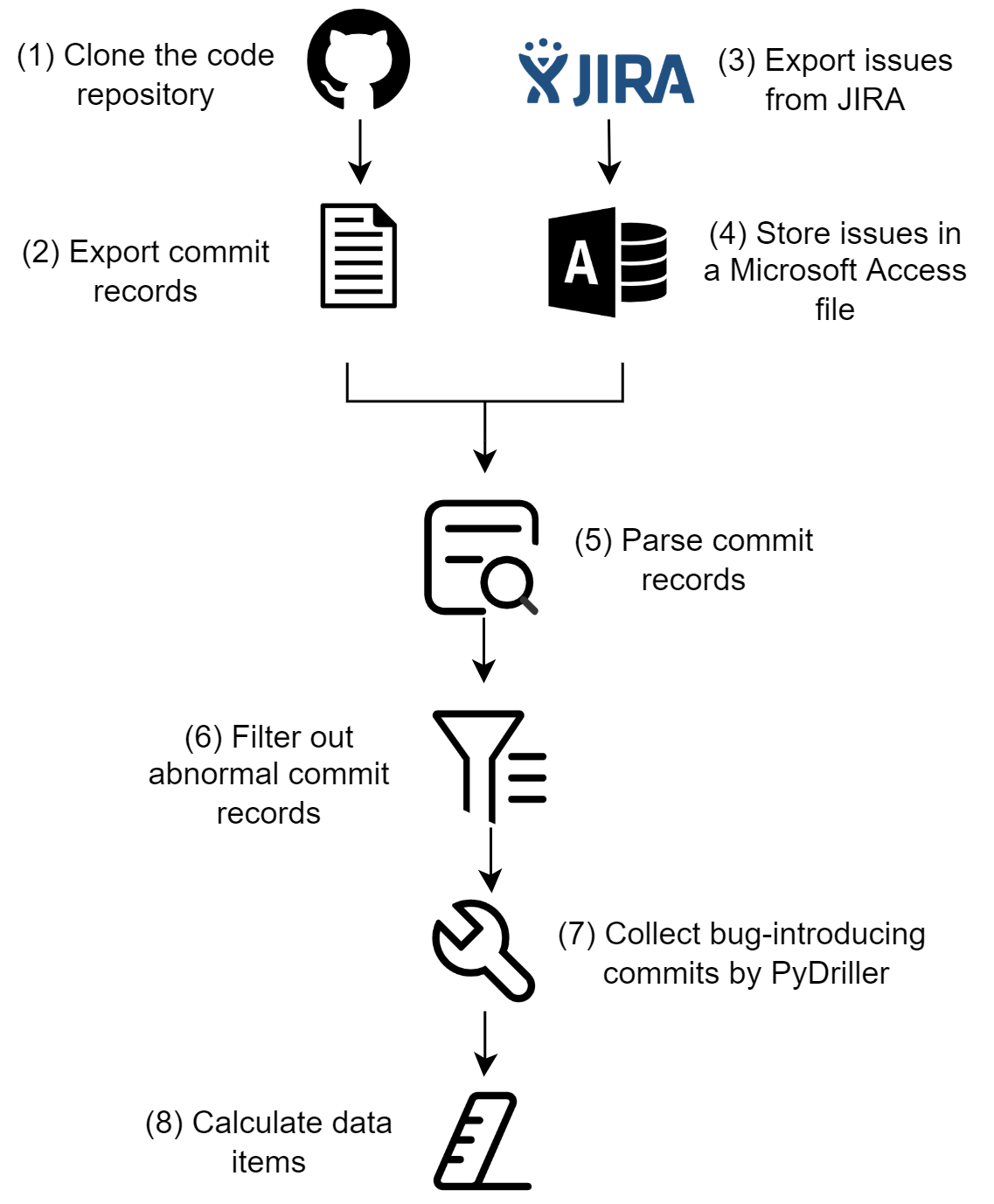}}
\caption{Procedure of data collection.}
\label{fig1}
\end{figure}

The data collection procedure for each selected project consists of seven steps (shown in Figure \ref{fig1}). The details of the steps are described as follows.

\noindent \textbf{Step 1}: Clone the repository of the project using TortoiseGit.

\noindent \textbf{Step 2}: Extract commit records from the repository to a text file for further parsing. In this step, we only exported the commit records of the master branch and the commit records merged to the master branch, but excluded the commit records corresponding to the MERGE operations of Git. A commit record corresponding to the MERGE operation in Git is duplicate with the merged commit records, in the sense that the file changes in the MERGE commit record are the same as the file changes in the merged commit records. In addition, the committer of the MERGE commit record is different from the committers of the merged commit records. Thus, the MERGE commit record should not be included to prevent double counting code changes.

\noindent \textbf{Step 3}: Export the issue reports. We manually exported all issues of the project from JIRA – deployed by the Apache Software Foundation.
%\footnote{\href{https://issues.apache.org/jira}{https://issues.apache.org/jira}}

\noindent \textbf{Step 4}: Store the exported issues in Step 3 to a Microsoft Access file, which is to facilitate data queries in the next steps.

\noindent \textbf{Step 5}: Parse commit records. If a commit is conducted to fix an issue, the committer would mention the issue ID explicitly in the commit message. The path and the number of lines of code changed of each modified source file can also be obtained in the commit record.

\noindent \textbf{Step 6}: Filter out abnormal commit records. Some commits contain changed source files with a large number of modified lines of code. For instance, if the involved source files are generated automatically, such files may be modified with tens of thousands of lines of code. Such commits should be filtered out. In this step, we filtered out commits in which more than 10,000 lines of code were modified.

\noindent \textbf{Step 7}: Collect bug-introducing commits by PyDriller. We use the PyDriller tool to identify bug-introducing commits based on the information of bug-fixing commits.

\noindent \textbf{Step 8}: Calculate data items. We calculated the data items listed in Table \ref{table:DataItemsForCommit} and Table \ref{table:DataItemsForFile}.

\subsection{Data Analysis}
To answer RQ1, besides the descriptive statistics, we also needed to calculate the entropy of PL use for each selected project. Suppose that the PLs used in a project are  $\{pl_1,pl_2,\cdots,pl_n\}$, and $k_i$ source files are written in PL $pl_i\left(1\leq i\leq n\right)$. Let 
\begin{equation}
p_i = k_i/\sum_{i=1}^nk_i. 
\end{equation}
\noindent Then, the entropy of PL use is
\begin{equation}
E_{PL}(n) = -\sum_{i=1}^np_ilog_2 p_i .
\end{equation}
\noindent Since the number of source files differs between different projects, we need to normalize the entropy of PL use to be comparable. The normalized entropy of PL use is 
\begin{equation}
EntropyPL(n) =
\begin{cases}
E_{PL}(n)/log_2 n &n>1 ,\\
0 &n=1 .
\end{cases}
\end{equation}

\noindent To answer RQ2, in addition to descriptive statistics, we conducted a Spearman test to calculate the correlation coefficient between the entropy of PL use and the proportion of MPLCs for all the selected projects. The test is significant at \emph{p-value} \textless{} 0.05, which means that the two tested variables have a significant correlation. The answer to RQ3 can be obtained by descriptive statistics. 

To answer RQ4, RQ5, RQ7, and RQ8, in addition to descriptive statistics, we performed Mann-Whitney U tests \citep{Fi2013} to examine if two groups are significantly different from each other. 
Both the Mann-Whitney U test and t-test can be used to test if there is a significant difference between two groups of instances of a variable. The Mann-Whitney U test is a non-parametric test, i.e., it does not make the assumption of a normal distribution of each group. In contrast, the t-test is a parametric test, i.e., it requires that the groups follow normal distributions.
Since the data of the variables to be tested do not necessarily follow a specific distribution, it is reasonable to use the Mann-Whitney U test -- a non-parametric test -- in this study. Besides, we measure the effect sizes for a Mann-Whitney U test by Pearson’s correlation coefficient \textit{r} \citep{Fi2013}. The test is significant at \emph{p-value} \textless{} 0.05, which means that the tested groups have a significant difference. 

To answer RQ6 and RQ8, we ran a Chi-squared test for each project \citep{Fi2013}. Specifically, for RQ6, the two variables of a Chi-squared test are: whether a fixed issue was reopened or not and whether it is connected to an MPLC or not; for RQ8, the two variables of a Chi-squared test are: MPLC or not, and bug-introducing or not. The effect size of a Chi-squared test is measured by the odds ratio (OR) \citep{Fi2013}. The test is significant at \emph{p-value} \textless{} 0.05, which means that the two variables are connected. 

Due to the multiple testing problem, we needed to perform the Benjamini-Hochberg procedure \citep{BeHo1995} to correct the \emph{p-values} of the Mann-Whitney U tests and Chi-squared tests.

\section{Study Results}
\label{chap:study}

We chose to examine around 500 Apache projects from JIRA, and all the projects meet project selection criterion C1. Then, we applied project selection criteria C2 - C6, and 30 projects were left. During this step, projects such as \textit{Tomcat} (excluded by C2), \textit{Hadoop} (excluded by C3), \textit{Lucy} (excluded by C4), \textit{HAWQ} (excluded by C5), and \textit{MADlib} (excluded by C6) were excluded. Finally, we applied criterion C7, 18 non-trivial Apache MPL OSS projects were selected for data collection. During this step, projects such as \textit{Avro} and \textit{Usergrid} were excluded since they have less than 100 MPLCs.

We collected data items described in Table \ref{table:DataItemsForCommit} and Table \ref{table:DataItemsForFile} from the 18 selected projects. The data of the selected projects were collected around the beginning of December of 2019. The study replication package is available online \citep{studymaterial}, including the collected dataset and a README file explaining the data fields in the dataset and how to use it. Specifically, the dataset includes the commit records exported from repositories and issues exported from JIRA of the 18 selected projects. Table \ref{table:DemographicInformation} shows the demographic information of the selected projects. The mean age of the projects is 9 years, the mean number of lines of code is 659K, the mean number of commits is 15,657, the mean number of contributors is 320, the mean number of issues is 8,645, and the mean number of bugs is 4,626. The number of PLs used in the projects ranges from 4 to 14, and the mean number of PLs used is 8. The percentage of code in the main PL (i.e., \%Main PL) of the projects ranges from 33.7\% to 74.7\%, and the mean percentage is 60.7\%. 
In addition, the projects belong to 10 domains, among which big data is the most popular domain.
In the rest of this section, we will present the results for each RQ.

\begin{table*}[]
\caption{Demographic information of the selected projects.}
\centering
\scalebox{0.95}{
\begin{tabular}{|c|r|r|r|r|r|r|r|c|r|c|}
\hline
\textbf{Project} & \multicolumn{1}{c|}{\textbf{Age (yr)}} & \multicolumn{1}{c|}{\textbf{\#LOC}} & \multicolumn{1}{c|}{\textbf{\#Commit}} & \multicolumn{1}{c|}{\textbf{\#Ctbtr}} & \multicolumn{1}{c|}{\textbf{\#Issue}} & \multicolumn{1}{c|}{\textbf{\#Bug}} & \multicolumn{1}{c|}{\textbf{\#PL}} & \textbf{Main PL} & \multicolumn{1}{c|}{\textbf{\%Main PL}} & \multicolumn{1}{c|}{\textbf{Domain}}\\ \hline
Airavata      & 9                                      & 1,057K                              & 9,290                                  & 45                                          & 3,364                                 & 1,535                               & 5                                  & Java             & 74.7                   & Distributed computing                \\ \hline
Ambari        & 9                                      & 1,093K                              & 24,588                                 & 134                                         & 25,261                                & 17,881                              & 11                                 & Java             & 45.9                  & Big data                 \\ \hline
Arrow         & 5                                      & 635K                                & 7,575                                  & 476                                         & 10,058                                & 3,410                               & 9                                 & C/C++            & 45.0                    & Big data               \\ \hline
Beam          & 5                                      & 1,053K                              & 29,021                                 & 676                                         & 11,019                                & 4,608                               & 8                                  & Java             & 72.3                   & Big data                \\ \hline
CarbonData    & 4                                      & 321K                                & 4,705                                  & 171                                         & 4,019                                 & 2,199                               & 5                                  & Scala            & 57.8                   & Big data                \\ \hline
CloudStack    & 10                                     & 912K                                & 32,644                                 & 329                                         & 10,312                                & 7,854                               & 7                                  & Java             & 59.5                   & Cloud computing               \\ \hline
CouchDB       & 12                                     & 123K                                & 12,376                                 & 162                                         & 3,292                                 & 1,878                               & 6                                  & Erlang           & 68.0                   & Database                \\ \hline
Dispatch      & 6                                      & 117K                                & 2,769                                  & 23                                          & 1,800                                 & 1,080                               & 5                                  & Python           & 42.3                   & Message router                \\ \hline
Ignite        & 6                                      & 2,056K                              & 27,056                                 & 241                                         & 12,727                                & 5,575                               & 8                                  & Java             & 74.6                   & Database                \\ \hline
Impala        & 9                                      & 640K                                & 9,429                                  & 146                                         & 10,168                                & 5,630                               & 5                                 & C/C++            & 54.5                    & Database               \\ \hline
Kafka         & 9                                      & 653K                                & 7,990                                  & 702                                         & 10,551                                & 5,809                               & 10                                 & Java             & 73.2                   & Event management                \\ \hline
Kylin         & 6                                      & 284K                                & 8,404                                  & 177                                         & 4,120                                 & 2,115                               & 6                                  & Java             & 71.3                   & Big data                \\ \hline
Ranger        & 6                                      & 327K                                & 3,371                                  & 77                                          & 3,013                                 & 2,048                               & 4                                  & Java             & 68.9                   & Data security                \\ \hline
Reef          & 8                                      & 283K                                & 3,873                                  & 68                                          & 2,063                                 & 498                                 & 6                                  & Java             & 52.9                   & Distributed computing                \\ \hline
Spark         & 10                                     & 973K                                & 28,142                                 & 1,634                                       & 28,983                                & 12,402                              & 6                                  & Scala            & 73.6                   & Big data                \\ \hline
Subversion    & 20                                     & 860K                                & 59,809                                 & 28                                          & 4,503                                 & 3,233                               & 6                                  & C/C++            & 65.8                   & Version control                \\ \hline
Thrift        & 14                                     & 258K                                & 6,101                                  & 340                                         & 5,283                                 & 2,923                               & 14                                 & C/C++            & 33.7                   & RPC framework                \\ \hline
Zeppelin      & 7                                      & 222K                                & 4,675                                  & 327                                         & 5,068                                 & 2,581                               & 6                                  & Java             & 59.3                   & Web notebook                \\ \hline
\textbf{Mean}          & \textbf{9}                                      & \textbf{659K}                                & \textbf{15,657}                                 & \textbf{320}                                         & \textbf{8,645}                                 & \textbf{4,626}                               & \textbf{8}                                  & \textbf{-}                & \textbf{60.7}         & \textbf{-}                          \\ \hline
\end{tabular}
}
\label{table:DemographicInformation}
\end{table*}

\subsection{Distribution of PLs Used in the Selected Projects (RQ1)}
The distribution of PLs used in the selected projects is shown in Table \ref{table:LanguagesInProjects}, where \textbf{X} denotes that the PL in the corresponding column is the main PL of the project in the corresponding row, \emph{\#Project} denotes the number of projects using the PL in the corresponding column, \emph{\#Project\textbf{X}} denotes the number of projects using the PL in the corresponding column as the main PL, and \emph{\#PL} denotes the number of PLs used by the project in the corresponding row. As we can see from this table, Java and Python are used in all the 18 projects; C/C++, JavaScript, and Ruby are also used in most projects; and Java is the main PL of 10 projects, followed by C/C++ which is the main PL of 4 projects. Besides, all projects are written in 4 or more PLs, with a maximum of 14 PLs for project \emph{Thrift}, followed by project \emph{Ambari} in 11 PLs. Furthermore, most (13 out 18, 72.2\%) of the projects use 5-8 PLs, and the most common PL combination consists of all the four PLs, i.e., C/C++, Java, JavaScript, and Python.

As shown in Table \ref{table:LanguagesInProjects}, we also calculated the entropy of PL use (i.e., EntropyPL in the last column of the table) for the selected MPL projects. Among all the selected MPL projects, \emph{Thrift} gets the largest EntropyPL of 0.793, with the most balanced use of PLs; and \emph{Kylin} has the smallest EntropyPL of 0.119, with the most unbalanced use of PLs. 

\begin{table*}[]
\caption{Programming languages used in the selected projects (RQ1).}
\centering
\begin{tabular}{|c|c|c|c|c|c|c|c|c|c|c|c|c|c|c|c|c|c|c|c|c|}
\hline
\textbf{Project} & \textbf{\rotatebox{270}{C/C++}} & \textbf{\rotatebox{270}{C\#}} & \textbf{\rotatebox{270}{Clojure}} & \textbf{\rotatebox{270}{CoffeeScript}} & \textbf{\rotatebox{270}{Erlang}} & \textbf{\rotatebox{270}{Go}} & \textbf{\rotatebox{270}{Haskell}} & \textbf{\rotatebox{270}{Java}} & \textbf{\rotatebox{270}{JavaScript}} & \textbf{\rotatebox{270}{Kotlin}} & \textbf{\rotatebox{270}{Objective-C}} & \textbf{\rotatebox{270}{Perl}} & \textbf{\rotatebox{270}{PHP}} & \textbf{\rotatebox{270}{Python}} & \textbf{\rotatebox{270}{Ruby}} & \textbf{\rotatebox{270}{Scala}} & \textbf{\rotatebox{270}{Swift}} & \textbf{\rotatebox{270}{TypeScript}} & \textbf{\rotatebox{270}{\#PL}} & \textbf{\rotatebox{270}{EntropyPL}} \\ \hline
Airavata      & x              &              &                  &                       &                 &             &                  & \textbf{X}    &                     &                 &                      &               & x            & x               &               &                &                & x                   & 5             & 0.237\\ \hline
Ambari        & x              & x            &                  & x                     &                 &             &                  & \textbf{X}    & x                   &                 &                      & x             & x            & x               & x             & x              &                & x                   & 11            & 0.525\\ \hline
Arrow         & \textbf{X}     & x            &                  &                       &                 & x           &                  & x             & x                   &                 & x                    &               &              & x               & x             &                &                & x                   & 9             & 0.648 \\ \hline
%\red{Avro}          & x              & x            &                  &                       &                 &             &                  & \textbf{X}    & x                   &                 &                      & x             & x            & x               & x             &                &                &                     & 8             & 0.636 \\ \hline
Beam          & x              &              &                  &                       &                 & x           &                  & \textbf{X}    & x                   & x               &                      &               &              & x               & x             &                &                & x                   & 8             & 0.295 \\ \hline
CarbonData    & x              & x            &                  &                       &                 &             &                  & x             &                     &                 &                      &               &              & x               &               & \textbf{X}     &                &                     & 5             & 0.444 \\ \hline
CloudStack    & x              & x            &                  &                       &                 &             &                  & \textbf{X}    & x                   &                 &                      & x             &              & x               & x             &                &                &                     & 7             & 0.209 \\ \hline
CouchDB       & x              &              &                  &                       & \textbf{X}      &             &                  & x             & x                   &                 &                      &               &              & x               & x             &                &                &                     & 6             &0.501 \\ \hline
Dispatch      & x              &              &                  &                       &                 &             &                  & x             & x                   &                 &                      &               &              & \textbf{X}      &               &                &                & x                   & 5             & 0.683 \\ \hline
Ignite        & x              & x            &                  &                       &                 &             &                  & \textbf{X}    & x                   &                 &                      &               & x            & x               &               & x              &                & x                   & 8             & 0.288 \\ \hline
Impala        & \textbf{X}     &              &                  &                       &                 &             &                  & x             & x                   &                 &                      &               &              & x               & x             &                &                &                     & 5             & 0.633\\ \hline
Kafka         & x              & x            & x                &                       &                 & x           &                  & \textbf{X}    & x                   &                 &                      &               & x            & x               & x             & x              &                &                     & 10            & 0.384 \\ \hline
Kylin         & x              &              &                  & x                     &                 &             &                  & \textbf{X}    & x                   &                 &                      &               &              & x               &               & x              &                &                     & 6             & 0.119 \\ \hline
Ranger        & x              &              &                  &                       &                 &             &                  & \textbf{X}    & x                   &                 &                      &               &              & x               &               &                &                &                     & 4             & 0.321 \\ \hline
Reef          & x              & x            &                  &                       &                 &             &                  & \textbf{X}    & x                   &                 &                      &               &              & x               &               & x              &                &                     & 6             & 0.394 \\ \hline
Spark         & x              &              &                  &                       &                 &             &                  & x             & x                   &                 &                      &               &              & x               & x             & \textbf{X}     &                &                     & 6             & 0.396 \\ \hline
Subversion    & \textbf{X}     & x            &                  &                       &                 &             &                  & x             &                     &                 &                      & x             &              & x               & x             &                &                &                     & 6             & 0.544 \\ \hline
Thrift        & \textbf{X}     & x            &                  &                       & x               & x           & x                & x             & x                   &                 & x                    & x             & x            & x               & x             &                & x              & x                   & 14            & 0.793 \\ \hline
%\red{Usergrid}      & x              & x            &                  &                       &                 &             &                  & \textbf{X}    & x                   &                 & x                    & x             & x            & x               & x             & x              & x              &                     & 11            & 0.313 \\ \hline
Zeppelin      &                &              &                  & x                     &                 &             &                  & \textbf{X}    & x                   &                 &                      &               &              & x               &               & x              &                & x                   & 6             & 0.548 \\ \hline
\#Project     & 17             & 9           & 1                & 3                     & 2               & 4           & 1                & 18            & 15                  & 1               & 2                    & 4             & 5            & 18              & 10            & 8              & 1              & 8                   & -             & - \\ \hline
\#Project\textbf{X}    & 4              & 0            & 0                & 0                     & 1               & 0           & 0                & 10            & 0                   & 0               & 0                    & 0             & 0            & 1               & 0             & 2              & 0              & 0                   & -             & - \\ \hline
\end{tabular}
\label{table:LanguagesInProjects}
\end{table*}

\subsection{Proportion of MPLCs in the Selected Projects (RQ2)}
As shown in Table \ref{table:PercentageOfMPLCs}, the total number of commits of all projects is 197,566, the total number of MPLCs is 18,372, and thus \textbf{the percentage of MPLCs is 9.3\% when taking all projects as a whole}. Please note that the number of commits for each project in Table \ref{table:PercentageOfMPLCs} is different from the number in Table \ref{table:DemographicInformation}. This is because we only considered the commits of and the commits merged into the master branch of the project repository. 
As presented in Table \ref{table:PercentageOfMPLCs}, the proportion of MPLCs over all commits of each selected OSS project ranges from 1.7\% to 41.0\%. 
%In projects \emph{Avro} (P4) and \emph{Usegrid} (P19), the number of MPLCs is even only around 50, much less than the other projects.

\begin{table}[]
\caption{Percentage of MPLCs in the selected projects (RQ2).}
\centering
\begin{tabular}{|c|r|r|r|}
\hline
\textbf{Project}   & \multicolumn{1}{c|}{\textbf{\#Commit}} & \multicolumn{1}{c|}{\textbf{\#MPLC}} & \multicolumn{1}{c|}{\textbf{\%MPLC}} \\ \hline
Airavata        & 5,562                                  & 269                                  & 4.8                                  \\ \hline
Ambari          & 19,667                                 & 1,491                                & 7.6                                  \\ \hline
Arrow           & 4,947                                  & 1,073                                & 21.7                                 \\ \hline
%\red{Avro}            & 1,637                                  & 52                                   & 3.2                                  \\ \hline
Beam            & 14,875                                 & 251                                  & 1.7                                  \\ \hline
CarbonData      & 3,330                                  & 1,366                                & 41.0                                 \\ \hline
CloudStack      & 23,285                                 & 1,786                                & 7.7                                  \\ \hline
CouchDB         & 7,114                                  & 560                                  & 7.9                                  \\ \hline
Dispatch        & 2,107                                  & 381                                  & 18.1                                 \\ \hline
Ignite          & 17,565                                 & 888                                  & 5.1                                  \\ \hline
Impala          & 7,485                                  & 1,975                                & 26.4                                 \\ \hline
Kafka           & 6,915                                  & 1,029                                & 14.9                                 \\ \hline
Kylin           & 5,587                                  & 126                                  & 2.3                                  \\ \hline
Ranger          & 2,445                                  & 203                                  & 8.3                                  \\ \hline
Reef            & 2,520                                  & 163                                  & 6.5                                  \\ \hline
Spark           & 21,782                                 & 2,772                                & 12.7                                 \\ \hline
Subversion      & 44,993                                 & 3,113                                & 6.9                                  \\ \hline
Thrift          & 4,409                                  & 512                                  & 11.6                                 \\ \hline
%\red{Usergrid}        & 6,791                                  & 45                                   & 0.7                                  \\ \hline
Zeppelin        & 2,978                                  & 414                                  & 13.9                                 \\ \hline
\textbf{Total} & \textbf{197,566}                       & \textbf{18,372}                      & \textbf{9.3}                         \\ \hline
\end{tabular}
\label{table:PercentageOfMPLCs}
\end{table}

\begin{figure}
\centerline{\includegraphics[width=3.5in]{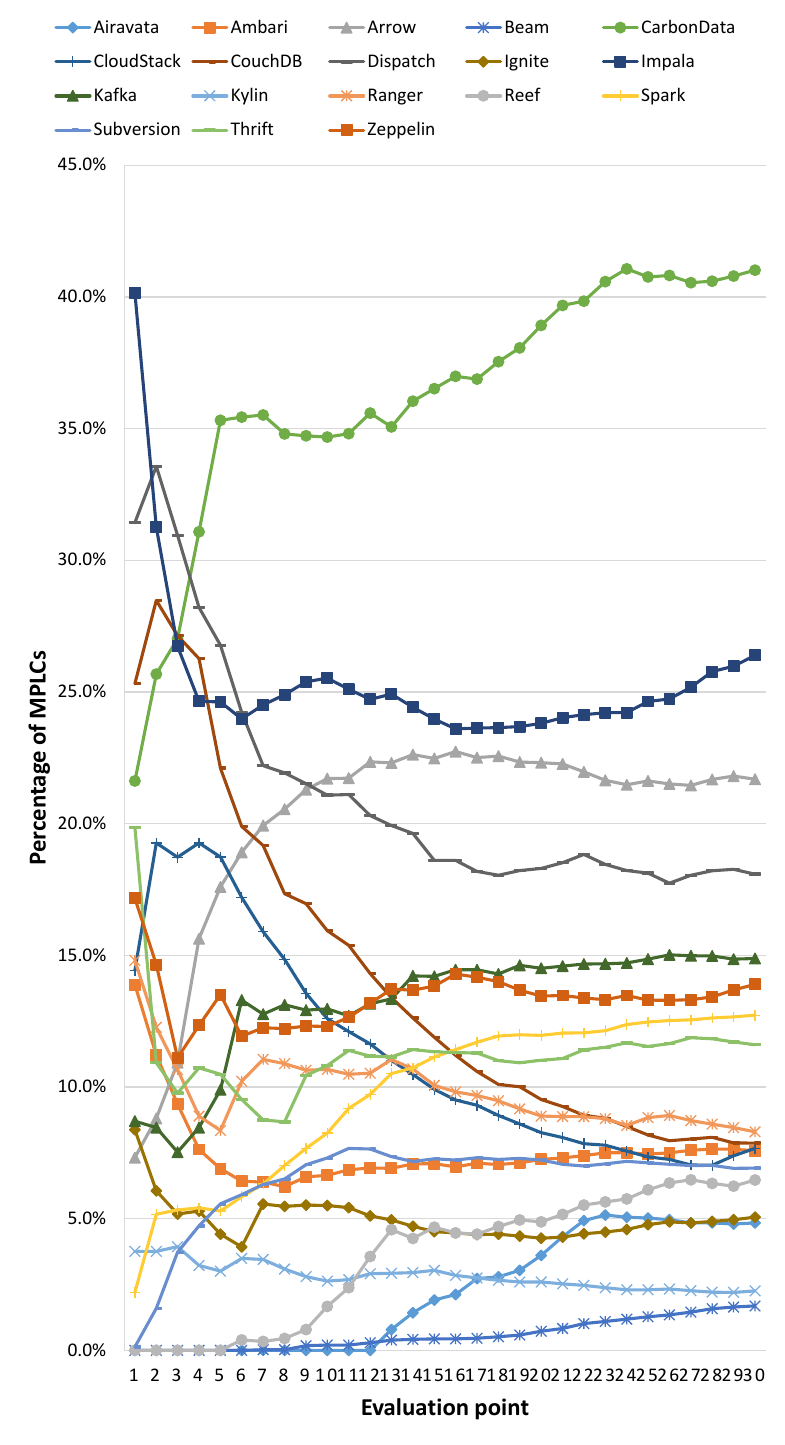}}
\caption{Trend of the proportion of MPLCs over time for each project (RQ2).}
\label{fig2}
\end{figure}

We further investigated the trend of the proportion of MPLCs in each selected project over time. To clearly display the trends of all the 18 projects in one diagram, we selected 30 evaluation points for each project to calculate the proportion of MPLCs. At the $k$th evaluation point, we calculated the proportion of MPLCs based on the first $k$ thirtieth of all commits of the project. As shown in Figure \ref{fig2},  \textbf{after relatively strong fluctuations in the early development stage of the projects, the proportion of MPLCs of most (15 out of 18, 83\%) of the projects tends to be stable in the late development stage}, while the proportion of MPLCs shows a long-term upward trend for \emph{CarbonData} and a long-term downward trend for \emph{CloudStack} and \emph{CouchDB}.

In addition, we conducted the Spearman test on the entropy of PL use (i.e., EntropyPL) and the proportion of MPLCs for all the selected projects. The resulting correlation coefficient is 0.653 and the \emph{p-value} is 0.003 (\textless{} 0.05), which suggests a significantly positive correlation. In other words, projects with higher proportions of MPLCs tend to be more balanced in terms of the use of PLs.

\subsection{Number of PLs Used in the Source Files Modified in MPLCs (RQ3)}
\begin{figure}
\centerline{\includegraphics[width=3.5in]{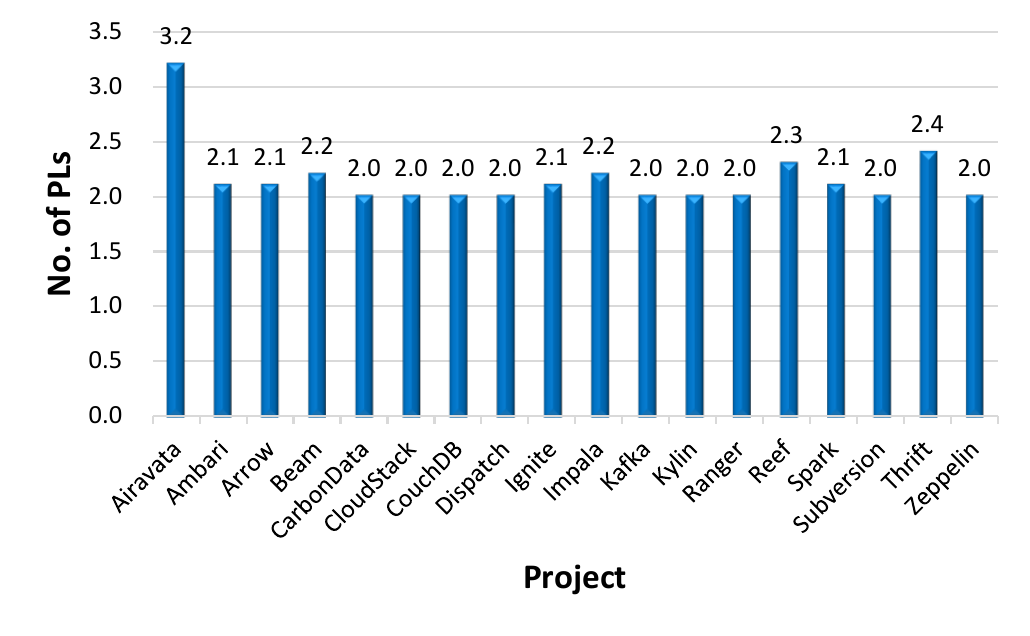}}
\caption{ Number of PLs used in the modified source files of MPLCs of the selected projects (RQ3).}
\label{fig3}
\end{figure}

Figure \ref{fig3} shows the average number of PLs used in the source files modified in MPLCs. Among the projects, on average, project \emph{Airavata} has 3.2 PLs used in the source files that are modified in each MPLC, and this project is the only project with no less than 3.0 PLs used in the source files modified in each MPLC. Most (15 out of 18, 83\%) of the projects have around 2.0 (i.e., 2.0-2.2) PLs for each MPLC on average. 
%Project \emph{Airavata} is the only project in which the source files modified in each MPLC are written in no less than 3.0 PLs on average. 
%project \emph{Avro} has the second most PLs of 2.5 for each MPLC,

We further explored how the number of PLs used distributes over MPLCs, and the results are shown in Table \ref{table:NumberOfPLs}. In this table, \emph{\#Ci} denotes the number of commits in which the modified source files are written in \emph{i} PLs, \emph{\%Ci} denotes the percentage of \emph{\#Ci} over \emph{\#MPLC}, and \emph{\#C5+} denotes the number of commits in which the modified source files are written in 5 or more PLs. As shown in Table \ref{table:NumberOfPLs}, taking all the projects as a whole, 91.7\%, 7.1\%, and 1.0\% of the MPLCs involve source files written in 2, 3, and 4 PLs, respectively; and only 0.2\% of the MPLCs involve source files written in 5 or more PLs. This means that \textbf{most MPLCs involve source files written in only 2 PLs, and it is not common for MPLCs to modify source files in more than 4 PLs.}

Table \ref{table:NumberOfPLs} also shows that most (14 out of 18, 78\%) of the projects do not have MPLCs with 5 or more PLs, and one third (6 out of 18) of the projects do not have MPLCs with more than 3 PLs.

\begin{table*}[]
\caption{Distribution of the number of PLs used in the modified source files of MPLCs of the selected projects (RQ3).}
\centering
\setlength{\tabcolsep}{3mm}{
\begin{tabular}{|c|r|r|r|r|r|r|r|r|r|}
\hline
\textbf{Project} & \multicolumn{1}{c|}{\textbf{\#MPLC}} & \multicolumn{1}{c|}{\textbf{\#C2}} & \multicolumn{1}{c|}{\textbf{\%C2}} & \multicolumn{1}{c|}{\textbf{\#C3}} & \multicolumn{1}{c|}{\textbf{\%C3}} & \multicolumn{1}{c|}{\textbf{\#C4}} & \multicolumn{1}{c|}{\textbf{\%C4}} & \multicolumn{1}{c|}{\textbf{\#C5+}} & \multicolumn{1}{c|}{\textbf{\%C5+}} \\ \hline
Airavata      & 269                                  & 59                                 & 21.9                               & 86                                 & 32.0                               & 124                                & 46.1                               & 0                                   & 0.0                                 \\ \hline
Ambari        & 1,491                                & 1,383                              & 92.8                               & 105                                & 7.0                                & 2                                  & 0.1                                & 1                                   & 0.1                                 \\ \hline
Arrow         & 1,073                                & 1,013                              & 94.4                               & 59                                 & 5.5                                & 1                                  & 0.1                                & 0                                   & 0.0                                 \\ \hline
%\red{Avro}          & 52                                   & 39                                 & 75.0                               & 9                                  & 17.3                               & 1                                  & 1.9                                & 3                                   & 5.8                                 \\ \hline
Beam          & 251                                  & 212                                & 84.5                               & 39                                 & 15.5                               & 0                                  & 0.0                                & 0                                   & 0.0                                 \\ \hline
CarbonData    & 1,366                                & 1,365                              & 99.9                               & 1                                  & 0.1                                & 0                                  & 0.0                                & 0                                   & 0.0                                 \\ \hline
CloudStack    & 1,786                                & 1,706                              & 95.5                               & 79                                 & 4.4                                & 1                                  & 0.1                                & 0                                   & 0.0                                 \\ \hline
CouchDB       & 560                                  & 547                                & 97.7                               & 10                                 & 1.8                                & 3                                  & 0.5                                & 0                                   & 0.0                                 \\ \hline
Dispatch      & 381                                  & 378                                & 99.2                               & 3                                  & 0.8                                & 0                                  & 0.0                                & 0                                   & 0.0                                 \\ \hline
Ignite        & 888                                  & 788                                & 88.7                               & 94                                 & 10.6                               & 5                                  & 0.6                                & 1                                   & 0.1                                 \\ \hline
Impala        & 1,975                                & 1,606                              & 81.3                               & 368                                & 18.6                               & 1                                  & 0.1                                & 0                                   & 0.0                                 \\ \hline
Kafka         & 1,029                                & 992                                & 96.4                               & 37                                 & 3.6                                & 0                                  & 0.0                                & 0                                   & 0.0                                 \\ \hline
Kylin         & 126                                  & 125                                & 99.2                               & 1                                  & 0.8                                & 0                                  & 0.0                                & 0                                   & 0.0                                 \\ \hline
Ranger        & 203                                  & 199                                & 98.0                               & 4                                  & 2.0                                & 0                                  & 0.0                                & 0                                   & 0.0                                 \\ \hline
Reef          & 163                                  & 115                                & 70.6                               & 47                                 & 28.8                               & 1                                  & 0.6                                & 0                                   & 0.0                                 \\ \hline
Spark         & 2,772                                & 2,491                              & 89.9                               & 277                                & 10.0                               & 4                                  & 0.1                                & 0                                   & 0.0                                 \\ \hline
Subversion    & 3,113                                & 3,039                              & 97.6                               & 53                                 & 1.7                                & 17                                 & 0.5                                & 4                                   & 0.1                                 \\ \hline
Thrift        & 512                                  & 436                                & 85.2                               & 31                                 & 6.1                                & 14                                 & 2.7                                & 31                                  & 6.1                                 \\ \hline
%\red{Usergrid}      & 45                                   & 43                                 & 95.6                               & 1                                  & 2.2                                & 0                                  & 0.0                                & 1                                   & 2.2                                 \\ \hline
Zeppelin      & 414                                  & 396                                & 95.7                               & 16                                 & 3.9                                & 2                                  & 0.5                                & 0                                   & 0.0                                 \\ \hline
\textbf{Total}    & \textbf{18,372}                      & \textbf{16,850}                    & \textbf{91.7}                      & \textbf{1,310}                     & \textbf{7.1}                       & \textbf{175}                       & \textbf{1.0}                       & \textbf{37}                         & \textbf{0.2}                        \\ \hline
\end{tabular}
}
\label{table:NumberOfPLs}
\end{table*}

\subsection{Change Complexity of MPLCs (RQ4)}
Change complexity can be measured by the number of lines of code modified (LOCM), number of source files modified (NOFM), number of directories modified (NODM), and entropy of source files modified (Entropy) \citep{LiLiLiMoLi2020}. In Table \ref{table:ChangeComplexity}, \emph{AveM} and \emph{AveN} denote the average value of the corresponding change complexity measure of MPLCs and non-MPLCs, respectively; \emph{\%Diff} denotes the percentage of the difference between \emph{AveM} and \emph{AveN}, i.e., 
\begin{equation}
\%Diff=\dfrac{AveM-AveN}{AveN}\times{100\%} . 
\end{equation}
As shown in Table \ref{table:ChangeComplexity}, for all the projects, on average, these four change complexity measures of MPLCs are much larger than those of non-MPLCs, respectively. Specifically, on average, the LOCM, NOFM, and NODM of MPLCs are larger than those of non-MPLCs by more than 100.0\% for most (15+ out of 18, 83\%+) of the projects, and the Entropy of MPLCs is larger than that of non-MPLCs by 42.2\% (i.e., project \textit{Arrow}) at least. We further ran Mann-Whitney U tests on the four measures of MPLCs and non-MPLCs for each project, and the corrected \emph{p-value} by the Benjamini–Hochberg procedure for each measure of each project is \textless{0.001} as shown in Table \ref{table:ChangeComplexity}. This indicates that all the four measures of MPLCs of each project are significantly larger than the measures for non-MPLCs, respectively. The effect size, i.e., ES(\textit{r}), for each Mann-Whitney U test is also shown in Table \ref{table:ChangeComplexity}. For each project, the effect size of NODM is larger than that of NOFM, and the latter is larger than the effect size of LOCM. For entropy, the effect sizes of 12 (66.7\%) projects are the smallest among the four change complexity measures,  the effect sizes of 5 (27.8\%) projects are the second smallest among the four change complexity measures, and only one project's effect size is the largest (slightly larger than the effect size of NODM for project \textit{CloudStack}) among the four change complexity measures. In this sense, among the four change complexity measures, \textbf{NODM has the strongest difference between MPLCs and non-MPLCs for the projects}. In summary, \textbf{the change complexity of MPLCs is significantly higher than that of non-MPLCs for each selected project.}

%\begin{table*}[]
\begin{sidewaystable}[!htp]
\caption{Change complexity of MPLCs and non-MPLCs (RQ4).}
\centering
\scalebox{0.9}{
\setlength{\tabcolsep}{1mm}{\begin{tabular}{|c|r|r|r|r|r|r|r|r|r|r|r|r|r|r|r|r|r|r|r|r|}
\hline
\multirow{2}{*}{\textbf{Project}} & \multicolumn{5}{c|}{\textbf{LOCM}}                                                                                                                    & \multicolumn{5}{c|}{\textbf{NOFM}}                                                                                                                    & \multicolumn{5}{c|}{\textbf{NODM}}                                                                                                                    & \multicolumn{5}{c|}{\textbf{Entropy}}                                                                                                                 \\ \cline{2-21} 
                                & \multicolumn{1}{c|}{\textbf{AveM}} & \multicolumn{1}{c|}{\textbf{AveN}} & \multicolumn{1}{c|}{\textbf{\%Diff}} & \multicolumn{1}{c|}{\textit{\textbf{p-value}}} & \multicolumn{1}{c|}{\textbf{ES(\textit{r})}} &
                                \multicolumn{1}{c|}{\textbf{AveM}} & \multicolumn{1}{c|}{\textbf{AveN}} & \multicolumn{1}{c|}{\textbf{\%Diff}} & \multicolumn{1}{c|}{\textit{\textbf{p-value}}} & \multicolumn{1}{c|}{\textbf{ES(\textit{r})}} &
                                \multicolumn{1}{c|}{\textbf{AveM}} & \multicolumn{1}{c|}{\textbf{AveN}} & \multicolumn{1}{c|}{\textbf{\%Diff}} & \multicolumn{1}{c|}{\textit{\textbf{p-value}}} & \multicolumn{1}{c|}{\textbf{ES(\textit{r})}} &
                                \multicolumn{1}{c|}{\textbf{AveM}} & \multicolumn{1}{c|}{\textbf{AveN}} & \multicolumn{1}{c|}{\textbf{\%Diff}} & \multicolumn{1}{c|}{\textit{\textbf{p-value}}} & \multicolumn{1}{c|}{\textbf{ES(\textit{r})}} \\ \hline
Airavata                       & 1,645                           & 313                             & 425.6                            & \textless{0.001}         & -0.266                      & 31                              & 5                               & 520.0                            & \textless{0.001}           & -0.296                    & 12                              & 3                               & 300.0                            & \textless{0.001}              & -0.320                 & 0.93                            & 0.50                            & 86.0                             & \textless{0.001}         & -0.199                      \\ \hline
Ambari                         & 541                             & 187                             & 189.3                            & \textless{0.001}          & -0.234                     & 13                              & 5                               & 160.0                            & \textless{0.001}          & -0.295                     & 7                               & 3                               & 133.3                            & \textless{0.001}       & -0.309                        & 0.88                            & 0.58                            & 51.7                             & \textless{0.001}        & -0.138                       \\ \hline
Arrow                          & 517                             & 262                             & 97.3                             & \textless{0.001}          & -0.253                     & 10                              & 6                               & 66.7                             & \textless{0.001}          & -0.344                     & 4                               & 2                               & 100.0                            & \textless{0.001}       & -0.453                        & 0.91                            & 0.64                            & 42.2                             & \textless{0.001}           & -0.152                    \\ \hline
Beam                           & 578                             & 226                             & 155.8                            & \textless{0.001}          & -0.078                     & 20                              & 6                               & 233.3                            & \textless{0.001}          & -0.128                     & 8                               & 3                               & 166.7                            & \textless{0.001}      & -0.156                        & 0.90                            & 0.57                            & 57.9                             & \textless{0.001}          & -0.060                     \\ \hline
CarbonData                     & 636                             & 264                             & 140.9                            & \textless{0.001}          & -0.394                     & 15                              & 5                               & 200.0                            & \textless{0.001}          & -0.536                     & 10                              & 3                               & 233.3                            & \textless{0.001}       & -0.588                        & 0.89                            & 0.59                            & 50.8                             & \textless{0.001}        & -0.217                       \\ \hline
CloudStack                     & 411                             & 156                             & 163.5                            & \textless{0.001}          & -0.207                     & 7                               & 4                               & 75.0                             & \textless{0.001}          & -0.273                     & 5                               & 2                               & 150.0                            & \textless{0.001}       & -0.314                        & 0.87                            & 0.34                            & 155.9                            & \textless{0.001}       & -0.323                        \\ \hline
CouchDB                        & 266                             & 95                              & 180.0                            & \textless{0.001}          & -0.238                     & 5                               & 2                               & 150.0                            & \textless{0.001}          & -0.402                     & 2                               & 1                               & 100.0                            & \textless{0.001}       & -0.608                        & 0.87                            & 0.27                            & 222.2                            & \textless{0.001}        & -0.357                       \\ \hline
Dispatch                       & 456                             & 145                             & 214.5                            & \textless{0.001}          & -0.396                     & 9                               & 3                               & 200.0                            & \textless{0.001}          & -0.510                     & 4                               & 2                               & 100.0                            & \textless{0.001}       & -0.625                        & 0.90                            & 0.42                            & 114.3                            & \textless{0.001}         & -0.328                      \\ \hline
Ignite                         & 1,019                           & 292                             & 249.0                            & \textless{0.001}          & -0.199                     & 40                              & 7                               & 471.4                            & \textless{0.001}          & -0.259                     & 16                              & 4                               & 300.0                            & \textless{0.001}       & -0.278                        & 0.90                            & 0.53                            & 69.8                             & \textless{0.001}       & -0.182                        \\ \hline
Impala                         & 455                             & 156                             & 191.7                            & \textless{0.001}          & -0.423                     & 12                              & 4                               & 200.0                            & \textless{0.001}          & -0.509                     & 5                               & 2                               & 150.0                            & \textless{0.001}       & -0.616                        & 0.90                            & 0.53                            & 69.8                             & \textless{0.001}       & -0.270                       \\ \hline
Kafka                          & 709                             & 202                             & 251.0                            & \textless{0.001}          & -0.351                     & 18                              & 5                               & 260.0                            & \textless{0.001}          & -0.431                     & 9                               & 3                               & 200.0                            & \textless{0.001}       & -0.484                        & 0.91                            & 0.57                            & 59.6                             & \textless{0.001}       & -0.172                        \\ \hline
Kylin                          & 274                             & 193                             & 42.0                             & \textless{0.001}          & -0.084                     & 7                               & 5                               & 40.0                             & \textless{0.001}          & -0.142                     & 5                               & 3                               & 66.7                             & \textless{0.001}       & -0.159                        & 0.86                            & 0.51                            & 68.6                             & \textless{0.001}        & -0.094                       \\ \hline
Ranger                         & 563                             & 262                             & 114.9                            & \textless{0.001}          & -0.188                     & 11                              & 5                               & 120.0                            & \textless{0.001}          & -0.289                     & 7                               & 3                               & 133.3                            & \textless{0.001}       & -0.336                        & 0.92                            & 0.51                            & 80.4                             & \textless{0.001}       & -0.189                        \\ \hline
Reef                           & 530                             & 254                             & 108.7                            & \textless{0.001}          & -0.188                     & 23                              & 8                               & 187.5                            & \textless{0.001}          & -0.256                     & 9                               & 4                               & 125.0                            & \textless{0.001}       & -0.275                        & 0.92                            & 0.61                            & 50.8                             & \textless{0.001}        & -0.104                       \\ \hline
Spark                          & 404                             & 126                             & 220.6                            & \textless{0.001}          & -0.289                     & 11                              & 4                               & 175.0                            & \textless{0.001}          & -0.364                     & 6                               & 3                               & 100.0                            & \textless{0.001}       & -0.388                        & 0.88                            & 0.54                            & 63.0                             & \textless{0.001}         & -0.224                      \\ \hline
Subversion                     & 235                             & 75                              & 213.3                            & \textless{0.001}          & -0.229                     & 7                               & 2                               & 250.0                            & \textless{0.001}          & -0.370                     & 3                               & 1                               & 200.0                            & \textless{0.001}       & -0.481                        & 0.86                            & 0.25                            & 244.0                            & \textless{0.001}       & -0.362                        \\ \hline
Thrift                         & 451                             & 128                             & 252.3                            & \textless{0.001}          & -0.290                     & 8                               & 3                               & 166.7                            & \textless{0.001}          & -0.423                     & 4                               & 1                               & 300.0                            & \textless{0.001}       & -0.541                        & 0.90                            & 0.38                            & 136.8                            & \textless{0.001}         & -0.299                      \\ \hline
Zeppelin                       & 541                             & 167                             & 224.0                            & \textless{0.001}          & -0.308                     & 10                              & 4                               & 150.0                            & \textless{0.001}          & -0.410                     & 6                               & 2                               & 200.0                            & \textless{0.001}       & -0.446                        & 0.90                            & 0.46                            & 95.7                             & \textless{0.001}        & -0.280                       \\ \hline
\end{tabular}}
\label{table:ChangeComplexity}}
\end{sidewaystable}
%\end{table*}

\subsection{Open time of issues fixed in MPLCs (RQ5)}

We studied the open time (i.e., the time from when an issue report is created to when the issue is resolved) of issues fixed in MPLCs, and the results are shown in Table \ref{table:OpenTimeOfIssues}, in which \emph{AveOTM} and \emph{AveOTN} denote the average open time of issues fixed in MPLCs and non-MPLCs respectively, \emph{\%Diff} denotes the difference between \emph{AveOTM} and \emph{AveOTN}, i.e.,
\begin{equation}
\%Diff=\dfrac{AveOTM-AveOTN}{AveOTN}\times{100\%} ,
\end{equation}
and \emph{p-value} denotes the corrected \emph{p-value} by the Benjamini-Hochberg procedure of the Mann-Whitney U test on the open time of issues fixed in MPLCs and non-MPLCs of a project. 

As shown in Table \ref{table:OpenTimeOfIssues}, 16 out of 18 (89\%) projects have significantly (\textit{p-value} \textless{0.05} with small effect sizes) longer open time of issues fixed in MPLCs than that of issues fixed in non-MPLCs; and the other 2 projects (i.e., \textit{CouchDB} and \textit{Subversion}, which corresponding cells of column \emph{p-value} are filled in grey) do not show a significant difference between the open time of issues fixed in MPLCs and that of issues fixed in non-MPLCs. The average open time of issues fixed in MPLCs is 8.0\% to 124.7\% longer than that of issues fixed in non-MPLCs for the 16 projects. It indicates that \textbf{issues fixed in MPLCs are likely to take longer to be resolved than issues fixed in non-MPLCs.}

\begin{table*}[]
\caption{Average open time of issues fixed in MPLCs of the selected projects (RQ5).}
\centering
\scalebox{1.0}{
\setlength{\tabcolsep}{1.5mm}{
\begin{tabular}{|c|r|r|r|r|r|r|}
\hline
\textbf{Project} & \begin{tabular}[c]{@{}c@{}}\textbf{AveOTM}\\ \textbf{(day)}\end{tabular} & \begin{tabular}[c]{@{}c@{}}\textbf{AveOTN}\\ \textbf{(day)}\end{tabular} & \multicolumn{1}{c|}{\textbf{\%Diff}} & \multicolumn{1}{c|}{\textit{\textbf{p-value}}} & \multicolumn{1}{c|}{\textbf{ES(\textit{r})}}\\ \hline
Airavata      & 95.22                                  & 58.95                                  & 61.5                              & 0.006            & -0.104           \\ \hline
Ambari        & 13.52                                  & 8.13                                   & 66.3                              & \textless{}0.001 & -0.139                       \\ \hline
Arrow         & 55.83                                  & 38.94                                  & 43.4                              & \textless{}0.001 & -0.113                                          \\ \hline
Beam          & 96.08                                  & 87.59                                  & 9.7                               & 0.008            & -0.053             \\ \hline
CarbonData    & 24.07                                  & 22.28                                  & 8.0                               & \textless{}0.001 & -0.120                        \\ \hline
CloudStack    & 113.71                                 & 50.60                                  & 124.7                             & \textless{}0.001  & -0.155                           \\ \hline
CouchDB       & 165.91                                 & 154.39                                 & 7.5                               & \cellcolor[HTML]{C0C0C0}0.792           & -0.014 \\ \hline
Dispatch      & 55.82                                  & 34.30                                  & 62.7                              & \textless{}0.001 & -0.168                       \\ \hline
Ignite        & 71.11                                  & 56.03                                  & 26.9                              & \textless{}0.001 & -0.053                       \\ \hline
Impala        & 160.51                                 & 78.66                                  & 104.1                             & \textless{}0.001  & -0.241                           \\ \hline
Kafka         & 98.26                                  & 67.84                                  & 44.8                              & \textless{}0.001 & -0.151                       \\ \hline
Kylin         & 123.40                                 & 61.64                                  & 100.2                             & 0.007             & -0.065                \\ \hline
Ranger        & 37.05                                  & 21.44                                  & 72.8                              & \textless{}0.001 & -0.137                       \\ \hline
Reef          & 41.99                                  & 29.24                                  & 43.6                              & \textless{}0.001 & -0.115                       \\ \hline
Spark         & 59.55                                  & 40.73                                  & 46.2                              & \textless{}0.001 & -0.111                       \\ \hline
Subversion    & 711.54                                 & 704.09                                 & 1.1                               & \cellcolor[HTML]{C0C0C0}0.411      & -0.022   \\ \hline
Thrift        & 159.95                                 & 119.91                                 & 33.4                              & \textless{}0.001 & -0.073                       \\ \hline
Zeppelin      & 48.30                                  & 35.64                                  & 35.5                              & \textless{}0.001 & -0.114                       \\ \hline
\end{tabular}
}}
\label{table:OpenTimeOfIssues}
\end{table*}

\subsection{Impact of MPLCs on issue reopen (RQ6)}
The proportions of reopened issues that were fixed in MPLCs and non-MPLCs are shown in Table \ref{table:ReopenedIssue}, in which \#RIMPLC and \#RINonMPLC denote the number of reopened issues that were fixed in MPLCs and non-MPLCs respectively, \#IMPLC and \#INonMPLC denote the number of issues that were fixed in MPLCs and non-MPLCs respectively, \%RIMPLC and \%RINonMPLC denote the percentage of \#RIMPLC over \#IMPLC and the percentage of \#RINonMPLC over \#INonMPLC respectively, \textit{p-value} denotes the corrected \textit{p-value} of the Chi-squared test and ES(OR) denotes the effect size of the test measured by the odds ratio.
As we can see from this table, the number of reopened issues that were fixed in MPLCs in each project is relatively small, from 0 to 98; and the proportion of the number of reopened issues fixed in MPLCs is larger than the proportion of that fixed in non-MPLCs in 8 out of 18 (44\%) projects. In addition, we conducted the Chi-squared test to investigate whether there is a connection between MPLCs and issue reopen in the selected projects. The \textit{p-values} for 89\% (16 out of 18) of the projects are greater than 0.05, which means there are no significant connections between MPLCs and issue reopen in these projects. For the other two projects, the test results show a significant connection between MPLCs and issue reopen (i.e., \textit{p-value} <0.05), and the effect sizes are medium. 
In summary, \textbf{there are no significant differences between MPLCs and non-MPLCs regarding the impact on issue reopen for 89\% of the projects}.

\begin{table*}[]
\caption{Reopened issues that were fixed in MPLCs and non-MPLCs (RQ6).}
\centering
\begin{tabular}{|c|r|r|r|r|r|r|r|r|}
\hline
\textbf{Project} & \textbf{\#RIMPLC} & \textbf{\#IMPLC} & \textbf{\%RIMPLC} & \textbf{\#RINonMPLC} & \textbf{\#INonMPLC} & \textbf{\%RINonMPLC} & \textbf{\textit{p-value}} & \textbf{ES(OR)}\\ \hline
Airavata         & 6                  & 61                & 9.8                & 47                    & 662                   & 7.1              & \cellcolor[HTML]{C0C0C0}0.600  &1.427    \\ \hline
Ambari           & 98                 & 1,325             & 7.4                & 814                   & 15,688                & 5.2              & \cellcolor[HTML]{C0C0C0}0.604  &1.080   \\ \hline
Arrow            & 10                 & 1,027             & 1.0                & 44                    & 3,318                 & 1.3              & \cellcolor[HTML]{C0C0C0}0.072     &0.291\\ \hline
Beam             & 2                  & 116               & 1.7                & 87                    & 2,649                 & 3.3              & \cellcolor[HTML]{C0C0C0}0.527   &0.516  \\ \hline
CarbonData       & 10                 & 814               & 1.2                & 14                    & 874                   & 1.6              & \cellcolor[HTML]{C0C0C0}0.435   &0.533  \\ \hline
CloudStack       & 11                 & 245               & 4.5                & 350                   & 3,637                 & 9.6              & 0.018     & 0.358\\ \hline
CouchDB          & 12                 & 118               & 10.2               & 20                    & 240                   & 8.3              & \cellcolor[HTML]{C0C0C0}0.643   & 0.800  \\ \hline
Dispatch         & 10                 & 252               & 4.0                & 32                    & 794                   & 4.0              & \cellcolor[HTML]{C0C0C0} 0.640   & 0.779  \\ \hline
Ignite           & 33                 & 470               & 7.0                & 218                   & 4,135                 & 5.3              & \cellcolor[HTML]{C0C0C0} 0.561    & 0.798  \\ \hline
Impala           & 59                 & 1,298             & 4.5                & 140                   & 2,768                 & 5.1              & \cellcolor[HTML]{C0C0C0} 0.061     & 0.658 \\ \hline
Kafka            & 30                 & 740               & 4.1                & 122                   & 3,155                 & 3.9              & \cellcolor[HTML]{C0C0C0} 0.063     & 0.514 \\ \hline
Kylin            & 2                  & 72                & 2.8                & 128                   & 1,717                 & 7.5              & \cellcolor[HTML]{C0C0C0} 0.402   & 0.354  \\ \hline
Ranger           & 10                 & 157               & 6.4                & 66                    & 1,449                 & 4.6              & \cellcolor[HTML]{C0C0C0} 0.556   & 1.425  \\ \hline
Reef             & 4                  & 97                & 4.1                & 29                    & 928                   & 3.1              & \cellcolor[HTML]{C0C0C0} 0.986   & 0.989  \\ \hline
Spark            & 94                 & 2,070             & 4.5                & 456                   & 11,110                & 4.1              & \textless{0.001}    & 0.507 \\ \hline
Subversion       & 0                  & 594               & 0.0                & 1                     & 1,011                 & 0.1              & \cellcolor[HTML]{C0C0C0} 0.570   & 0.000  \\ \hline
Thrift           & 45                 & 354               & 12.7               & 278                   & 2,087                 & 13.3             & \cellcolor[HTML]{C0C0C0} 0.520   & 0.805  \\ \hline
Zeppelin         & 2                  & 271               & 0.7                & 15                    & 1,414                 & 1.1              & \cellcolor[HTML]{C0C0C0} 0.564   & 0.345  \\ \hline
\end{tabular}
\label{table:ReopenedIssue}
\end{table*}

\subsection{Bug proneness of source files modified in MPLCs (RQ7)}
Bug proneness of a source file can be measured by defect density (DD) of the file, i.e., NOB/LOC of the file \citep{LiLiLi2017}. We calculated the defect density of the source files modified in MPLCs and that of the source files modified only in non-MPLCs. The results are shown in Table \ref{table:DefectDensity}, in which \emph{\#FileAll} denotes the number of all source files of a project, \emph{LOC} denotes the number of lines of code in a source file, \emph{Ave. LOC} denotes the average \emph{LOC} of all source files of a project, \emph{\#FileM} denotes the number of source files modified in MPLCs, \emph{\#FileN} denotes the number of source files modified only in non-MPLCs, \emph{DDM} denotes the defect density of a source file modified in MPLCs, \emph{DDN} denotes the defect density of a source file modified only in non-MPLCs, and \emph{\%Diff} denotes the percentage of the difference between \emph{Ave. DDM} and \emph{Ave. DDN}, i.e., 
\begin{equation}
\%Diff=\dfrac{Ave. DDM-Ave. DDN}{Ave. DDN}\times{100\%}.
\end{equation}
We ran the Mann-Whitney U test to compare the \emph{DDM} and \emph{DDN} of source files in each project and performed the Benjamini-Hochberg procedure for the results, and the corrected \emph{p-value} is shown in Table \ref{table:DefectDensity}. As we can see from the table, \emph{DDM} is significantly larger than \emph{DDN} with \emph{p-value} \textless{} 0.05 in 15 out of 18 (83\%) projects, the effect sizes are between -0.514 and -0.084 (most are small or medium), and the difference between \emph{DDM} and \emph{DDN} ranges from 17.2\% to 3,681.8\%; \emph{DDM} is significantly smaller than \emph{DDN} with \emph{p-value} \textless{} 0.05 in one project (\textit{Ignite}, marked with * according to the corresponding \emph{p-value} in Table \ref{table:DefectDensity}); and there is no significant difference between \emph{DDM} and \emph{DDN} in the other 2 projects (\textit{Airavata} and \textit{CloudStack}, with \emph{p-value} \textgreater{} 0.05, and the corresponding cells of column \emph{p-value} are filled in grey in Table \ref{table:DefectDensity}). In other words, \textbf{the defect density of source files that have been modified in MPLCs is likely larger than the defect density of source files that have never been modified in MPLCs.}

\begin{table*}[]
\caption{Defect density of source files modified in MPLCs and non-MPLCs (RQ7).}
\centering
\begin{tabular}{|c|r|r|r|r|r|r|r|r|r|}
\hline
\textbf{Project} & \multicolumn{1}{c|}{\textbf{\#FileAll}} & \multicolumn{1}{c|}{\textbf{Ave. LOC}} & \multicolumn{1}{c|}{\textbf{\#FileM}} & \multicolumn{1}{c|}{\textbf{\#FileN}} & \multicolumn{1}{c|}{\textbf{Ave. DDM}} & \multicolumn{1}{c|}{\textbf{Ave. DDN}} & \multicolumn{1}{c|}{\textbf{\%Diff}} & \multicolumn{1}{c|}{\textit{\textbf{p-value}}} & \multicolumn{1}{c|}{\textbf{ES(\textit{r})}} \\ \hline
Airavata      & 993                                       & 1,065                             & 450                                     & 543                                     & 0.00174                             & 0.00325                             & -46.5                             & \cellcolor[HTML]{C0C0C0}0.381            & -0.028      \\ \hline
Ambari        & 5,645                                     & 194                               & 2,272                                   & 3,373                                   & 0.03636                             & 0.01848                             & 96.8                              & \textless{0.001}               & -0.297                \\ \hline
Arrow         & 2,851                                     & 223                               & 1,757                                   & 1,094                                   & 0.01041                             & 0.00604                             & 72.4                              & \textless{0.001}               & -0.132                 \\ \hline
Beam          & 5,812                                     & 181                               & 2,344                                   & 3,468                                   & 0.01034                             & 0.00325                             & 218.2                             & \textless{0.001}               & -0.514                 \\ \hline
CarbonData    & 1,484                                     & 216                               & 1,244                                   & 240                                     & 0.00770                             & 0.00290                             & 165.5                             & \textless{0.001}               & -0.234                 \\ \hline
CloudStack    & 2,395                                     & 381                               & 1,346                                   & 1,049                                   & 0.00256                             & 0.00247                             & 3.6                               & \cellcolor[HTML]{C0C0C0}0.075            & -0.037       \\ \hline
CouchDB       & 557                                       & 221                               & 173                                     & 384                                     & 0.00038                             & 0.00009                             & 322.2                             & 0.008               & -0.114                            \\ \hline
Dispatch      & 332                                       & 351                               & 261                                     & 71                                      & 0.02112                             & 0.00517                             & 308.5                             & \textless{0.001}               & -0.410                 \\ \hline
Ignite        & 12,270                                    & 168                               & 4,924                                   & 7,346                                   & 0.01085                             & 0.01212                             & -10.5                             & \textless{0.001}*              & -0.084                 \\ \hline
Impala        & 2,364                                     & 271                               & 1,784                                   & 580                                     & 0.01549                             & \textless{0.001}86                             & 218.7                             & \textless{0.001}               & -0.410                 \\ \hline
Kafka         & 3,341                                     & 195                               & 2,365                                   & 976                                     & 0.01395                             & 0.01190                             & 17.2                              & \textless{0.001}               & -0.084                 \\ \hline
Kylin         & 1,831                                     & 155                               & 328                                     & 1,503                                   & 0.01165                             & 0.00784                             & 48.6                              & \textless{0.001}               & -0.125                 \\ \hline
Ranger        & 1,480                                     & 221                               & 579                                     & 901                                     & 0.01861                             & 0.01394                             & 33.5                              & \textless{0.001}               & -0.205                 \\ \hline
Reef          & 3,590                                     & 79                                & 499                                     & 3,091                                   & 0.01271                             & \textless{0.001}21                             & 201.9                             & \textless{0.001}               & -0.236                 \\ \hline
Spark         & 4,854                                     & 201                               & 3,759                                   & 1,095                                   & 0.01499                             & 0.00711                             & 110.8                             & \textless{0.001}               & -0.213                 \\ \hline
Subversion    & 1,571                                     & 548                               & 1,296                                   & 275                                     & \textless{0.001}16                             & 0.00011                             & 3,681.8                            & \textless{0.001}              & -0.270                  \\ \hline
Thrift        & 1,322                                     & 195                               & 839                                     & 483                                     & 0.01307                             & 0.00742                             & 76.1                              & \textless{0.001}               & -0.223                 \\ \hline
Zeppelin      & 1,197                                     & 185                               & 716                                     & 481                                     & 0.00945                             & \textless{0.001}45                             & 112.4                             & \textless{0.001}               & -0.258                 \\ \hline
\end{tabular}
\label{table:DefectDensity}
\end{table*}

\subsection{Bug introduction of MPLCs (RQ8)}
We studied the impact of MPLCs on software quality in terms of bug introduction, and the results are presented as follows. We first investigated the difference between the proportion of bug-introducing MPLCs over all MPLCs and the proportion of bug-introducing non-MPLCs over all non-MPLCs. The results are shown in Table \ref{table:PercentageBugInducingCommits}, where \#BIMPLC and \#BINonMPLC denote the number of bug-introducing MPLCs and the number of bug-introducing non-MPLCs respectively. As we can see from this table, the percentage of bug-introducing MPLCs (i.e., \%BIMPLC) for each project ranges from 23.6\% to 85.8\% and the average is 54.2\%; in contrast, the percentage of bug-introducing non-MPLCs (i.e., \%BINonMPLC) for each project falls into [3.4\%, 64.8\%] and the average is 32.1\%. 
In particular, the percentage of bug-introducing MPLCs is larger than the percentage of bug-introducing non-MPLCs in each project. 

We ran the Chi-squared test for each project to investigate if there is a connection between MPLCs and bugs, and the results are shown in Table \ref{table:PercentageBugInducingCommits}. There are significant connections between MPLCs and bugs for 89\% (i.e., 16 out of 18) of the projects, and the corresponding effect sizes are medium (\textgreater{1.500}) or strong (\textgreater{3.000}).
This means that MPLCs are significantly associated with bugs in 89\% of the projects. 
In other words, \textbf{MPLCs are more likely to introduce bugs than non-MPLCs in the selected projects.}

We further looked into the number of bugs introduced by MPLCs. The results are shown in Table \ref{table:BugInductionFocus}, where AveNumIBMPLC and AveNumIBNonMPLC denote the average number of introduced bugs over all bug-introducing MPLCs and all bug-introducing non-MPLCs respectively, and \textit{p-value} is the result of the Mann-Whitney U test for comparing AveNumIBMPLC and AveNumIBNonMPLC after the Benjamini-Hochberg procedure. As can be seen from this table, in 17 out of the 18 (94\%) projects, AveNumIBMPLC is significantly larger than AveNumIBNonMPLC with small effect sizes, and in the rest project \textit{Airavata}, there is no significant difference between AveNumIBMPLC and AveNumIBNonMPLC. In summary, \textbf{in 94\% of the projects, the average number of introduced bugs of all bug-introducing MPLCs is significantly larger than that of all bug-introducing non-MPLCs}.

\begin{table*}[]
\caption{Results of Chi-squared tests on MPLCs and bugs (RQ8).}
\centering
\begin{tabular}{|c|r|r|r|r|r|r|r|r|}
\hline
\textbf{Project} & \textbf{\#BIMPLC} & \textbf{\#MPLC} & \textbf{\%BIMPLC} & \textbf{\#BINonMPLC} & \textbf{\#NonMPLC} & \textbf{\%BINonMPLC} & \textbf{\textit{p-value}} & \textbf{ES(OR)}\\ \hline
Airavata         & 69                & 269             & 25.7              & 1,185                & 5,293     & 22.4         & \cellcolor[HTML]{C0C0C0}0.212            & 1.196     \\ \hline
Ambari           & 1,279             & 1,491           & 85.8              & 11,782               & 18,176    & 64.8         & \textless{0.001}     & 3.274            \\ \hline
Arrow            & 596               & 1,073           & 55.5              & 1,501                & 3,874     & 38.7         & \textless{0.001}     & 1.975            \\ \hline
Beam             & 104               & 251             & 41.4              & 3,086                & 14,624    & 21.1         & \textless{0.001}     & 2.645            \\ \hline
CarbonData       & 899               & 1,366           & 65.8              & 655                  & 1,964     & 33.4         & \textless{0.001}     & 3.847            \\ \hline
CloudStack       & 463               & 1,786           & 25.9              & 5,249                & 21,499    & 24.4         & \cellcolor[HTML]{C0C0C0}0.163            & 1.083     \\ \hline
CouchDB          & 132               & 560             & 23.6              & 221                  & 6,554     & 3.4          & \textless{0.001}     & 8.838            \\ \hline
Dispatch         & 279               & 381             & 73.2              & 766                  & 1,726     & 44.4         & \textless{0.001}     & 3.428            \\ \hline
Ignite           & 477               & 888             & 53.7              & 4,777                & 16,677    & 28.6         & \textless{0.001}     & 2.891            \\ \hline
Impala           & 1,411             & 1,975           & 71.4              & 2,464                & 5,510     & 44.7         & \textless{0.001}     & 3.093            \\ \hline
Kafka            & 723               & 1,029           & 70.3              & 2,658                & 5,886     & 45.2         & \textless{0.001}     & 2.869            \\ \hline
Kylin            & 65                & 126             & 51.6              & 1,640                & 5,461     & 30.0         & \textless{0.001}     & 2.483           \\ \hline
Ranger           & 165               & 203             & 81.3              & 1,112                & 2,242     & 49.6         & \textless{0.001}     & 4.412            \\ \hline
Reef             & 83                & 163             & 50.9              & 396                  & 2,357     & 16.8         & \textless{0.001}     & 5.138            \\ \hline
Spark            & 1,324             & 2,772           & 47.8              & 6,465                & 19,010    & 34.0         & \textless{0.001}     & 1.774            \\ \hline
Subversion       & 851               & 3,113           & 27.3              & 4,389                & 41,880    & 10.5         & \textless{0.001}     & 3.214            \\ \hline
Thrift           & 306               & 512             & 59.8              & 1,214                & 3,897     & 31.2         & \textless{0.001}     & 3.283            \\ \hline
Zeppelin         & 266               & 414             & 64.3              & 879                  & 2,564     & 34.3         & \textless{0.001}     & 3.445            \\ \hline
\textbf{Average}          & -        & -               & \textbf{54.2}     & -                    & -         & \textbf{32.1} & -        & -         \\ \hline
\end{tabular}
\label{table:PercentageBugInducingCommits}
\end{table*}

\begin{table*}[]
\caption{Bug introduction of MPLCs and non-MPLCs (RQ8).}
\centering
\begin{tabular}{|c|r|r|r|r|r|r|}
\hline
\textbf{Project} & \textbf{\#BIMPLC} & \textbf{AveNumIBMPLC} & \textbf{\#BINonMPLC} & \textbf{AveNumIBNonMPLC} & \textit{\textbf{p-value}}   & \textbf{ES(\textit{r})} \\ \hline
Airavata         & 69              & 1.42                & 1,185               & 1.54                    & {\cellcolor[HTML]{C0C0C0} 0.813}  & -0.007\\ \hline
Ambari           & 1,279           & 7.12                & 11,782              & 3.76                    & \textless{0.001}             & 0.174 \\ \hline
Arrow            & 596             & 3.64                & 1,501               & 2.31                    & \textless{0.001}             & 0.174 \\ \hline
Beam             & 104             & 2.16                & 3,086               & 1.62                    & \textless{0.001}             & 0.066 \\ \hline
CarbonData       & 899             & 3.80                & 655                 & 2.09                    & \textless{0.001}             & 0.304 \\ \hline
CloudStack       & 463             & 2.99                & 5,249               & 1.93                    & \textless{0.001}             & 0.118 \\ \hline
CouchDB          & 132             & 1.62                & 221                 & 1.40                    & 0.005             & 0.153            \\ \hline
Dispatch         & 279             & 3.99                & 766                 & 2.28                    & \textless{0.001}             & 0.252 \\ \hline
Ignite           & 477             & 3.33                & 4,777               & 2.18                    & \textless{0.001}             & 0.101 \\ \hline
Impala           & 1,411           & 4.27                & 2,464               & 2.52                    & \textless{0.001}             & 0.240 \\ \hline
Kafka            & 723             & 6.30                & 2,658               & 3.19                    & \textless{0.001}             & 0.262 \\ \hline
Kylin            & 65              & 2.57                & 1,640               & 1.78                    & \textless{0.001}             & 0.100 \\ \hline
Ranger           & 165             & 6.42                & 1,112               & 3.17                    & \textless{0.001}             & 0.213 \\ \hline
Reef             & 83              & 1.86                & 396                 & 1.66                    & 0.012             & 0.117           \\ \hline
Spark            & 1,324           & 3.53                & 6,465               & 2.36                    & \textless{0.001}             & 0.139 \\ \hline
Subversion       & 851             & 2.18                & 4,389               & 1.56                    & \textless{0.001}             & 0.118 \\ \hline
Thrift           & 306             & 3.10                & 1,214               & 1.94                    & \textless{0.001}             & 0.181 \\ \hline
Zeppelin         & 266             & 3.38                & 879                 & 1.91                    & \textless{0.001}             & 0.195 \\ \hline
\end{tabular}
\label{table:BugInductionFocus}
\end{table*}

\section{Discussion}
\label{chap:discus}

In this section, we interpret the results of the study according to the RQs and discuss the implications of the results for both practitioners and researchers. 
\subsection{Interpretation of Study Results}

\emph{\textbf{RQ1}}: Although different PLs are used in different projects, they tend to use a specific combination of PLs.
The entropy of PL use, i.e., EntropyPL, measures the degree to which the use of PLs is balanced. The result on the EntropyPL suggests that a higher degree of balanced use of PLs does not depend on a larger number of PLs used. For instance, the project \emph{Dispach} uses the second least PLs (i.e., 5 PLs) but gets the second largest EntropyPL of 0.683.

The number of PLs employed and the EntropyPL do not depend on the domain of the project, but on the functionalities provided by the project. For instance, both \textit{Arrow} and \textit{Kylin} belong to the domain of big data, but the EntropyPL of \textit{Arrow} is much larger than that of \textit{Kylin}; Both \textit{Ambari} and \textit{CarbonData} belong to the domain of big data, but the number of PLs used in \textit{Ambari} is much larger than that in \textit{CarbonData}. Although \textit{Arrow} and \textit{Thrift} belong to different domains, both provide libraries used by multiple PLs. \textit{Arrow} provides libraries that are ``available for C, C++, C\#, Go, Java, JavaScript, Julia, MATLAB, Python, R, Ruby, and Rust\footnote{https://arrow.apache.org}''; \textit{Thrift} is for scalable cross-language services development, and ''combines a software stack with a code generation engine to build services that work efficiently and seamlessly between C++, Java, Python, PHP, Ruby, Erlang, Perl, Haskell, C\#, Cocoa, JavaScript, Node.js, Smalltalk, OCaml and Delphi and other languages\footnote{https://thrift.apache.org}''. 

\emph{\textbf{RQ2}}: Only 9.3\% of the commits are MPLCs when taking all projects as a whole, and the other 90.7\% of the commits are non-MPLCs. It indicates that developers tend to make mono-language changes despite the MPL development. However, from the perspective of individual projects, the proportion of MPLCs may differ greatly. As we can see from Table  \ref{table:DemographicInformation} and Table \ref{table:PercentageOfMPLCs}, a greater number of PLs used and a lower percentage of code in the main PL do not necessarily mean a larger proportion of MPLCs in a project. One possible reason is that design quality may play an important role. For instance, higher modularity may reduce the likelihood of MPLCs.

Although the proportion of MPLCs of the selected projects differs from one to one, the proportion of MPLCs goes to a relatively stable level for most (83\%) of the selected projects. This is an interesting phenomenon, which may indicate certain balanced status in the development, e.g., stable architecture design. It is worth further investigation to explore what factors play dominant roles in this phenomenon.

The strong positive correlation between the proportion of MPLCs and the Entropy of the use of PLs indicates that the proportion of MPLCs is related to how balanced different PLs are used. On the other hand, a higher proportion of MPLCs may not depend on a larger number of PLs used. As we can see in Table \ref{table:LanguagesInProjects}, projects \textit{Beam}, \textit{CloudStack}, \textit{Ignite}, and \textit{Kafka} use a relatively large number of PLs, but they have only relatively small EntropyPL, smaller than most projects (e.g., \textit{Dispatch} and \textit{Impala}) with 5 or 6 PLs used.

\emph{\textbf{RQ3}}: Most of MPLCs involve source files written in two PLs, which may be a natural choice of developers. Source files in more PLs to be modified in a commit may lead to higher complexity of the code change, which requires more comprehensive consideration of the potential influence of the code change on the quality of the software systems. However, there is a lack of effective tools to automatically analyze change impact in an MPL context, and MPL code analysis still remains a rather challenging problem \citep{ShMiAb2019}.

\emph{\textbf{RQ4}}: Change complexity of MPLCs is significantly higher than that of non-MPLCs, which is not surprising. MPLCs involve source files written in multiple PLs, and source files in different PLs are usually distributed over different components. Therefore, MPLCs tend to exert relatively global impact on the software system, and the change complexity of MPLCs is likely to be higher. In addition, the results of all the four change complexity measures (i.e., LOCM, NOFM, NODM, and Entropy) are perfectly consistent, which increases the confidence in the finding that changes in MPLCs are more complex than changes in non-MPLCs. Finally, the information on whether a code change is an MPLC can facilitate effort estimation in project management. 

\emph{\textbf{RQ5}}: The results on the open time of issues show that issues fixed in MPLCs likely take longer to be resolved than issues fixed in non-MPLCs. The open time of issues generally depends on two factors: the priority of issues and the difficulty of resolving issues. We investigated whether there is a significant difference on the priority of issues fixed in MPLCs and non-MPLCs, and found that there is no significant difference between the priority of issues fixed in MPLCs and non-MPLCs for most (10/18) of the selected projects. The results on the issue priority are not presented in this paper due to derivation from the focus of this work, but have been available online\footnote{https://github.com/ASSMS/JSS/blob/main/IssuePriority.pdf}. Thus, the main reason for issues fixed in MPLCs taking longer to be resolved may be that such issues are more difficult to be fixed, which is evidenced by the results on RQ4. 

\emph{\textbf{RQ6}}: There is no significant connection between MPLCs and issue reopen. Although the change complexity of MPLCs are significantly higher than that of non-MPLCs, it does not mean that MPLCs will result in more rework of bug-fixing.

\emph{\textbf{RQ7}}: Source files modified in MPLCs are likely to be more bug-prone, thus, the proportion of source files that have been modified in MPLCs can be used as an indicator of risk of bug introduction in MPL software systems. Source files in various PLs modified in an MPLC indicate that these source files are linked together due to (in)direct dependencies. In addition, source files in different PLs are communicated through dedicated mechanisms (e.g., JNI) and there is a lack of cross-language analysis tools, which increases the difficulty of bug fixing and consequently results in higher bug proneness of the involved source files in MPLCs.

\emph{\textbf{RQ8}}: (1) One potential reason for that MPLCs are more likely to introduce bugs is that: an MPLC often involves changes in multiple modules or subsystems, which is more complex to analyze and understand. 
For the same reason, an bug-introducing MPLC tends to introduce more bugs than a bug-introducing non-MPLC. (2) As shown in Table \ref{table:BugInductionFocus}, an MPLC or non-MPLC can introduce a relatively large number (more than one and even seven) of bugs. The reason is that a bug may be introduced by multiple commits. Since we consider a bug-introducing commit of bug $B$ as the last commit containing one or more files modified in the commit in which bug $B$ is fixed, bug $B$ can have more than one bug-introducing commit.

\subsection{Implications for Practitioners}
\textbf{Prevent too many MPLCs by architecture design of MPL software systems.} Since the change complexity of MPLCs is significantly higher than non-MPLCs, too many MPLCs happening in an MPL system will result in a considerable increase of effort for making changes to the system. An MPLC tends to be at the architecture level in light of that multiple components are modified in the MPLC. Thus, it is wise to design a more maintainable architecture for an MPL system. 
When a relatively high proportion of commits are MPLCs in an MPL system, there is a necessity to assess the maintainability (especially modularity) of the architecture of this system \citep{MoCaKaXiFe2016,WoCaKiDa2011}, and then to improve the architecture design through e.g., refactoring.

\textbf{Pay special attention to source files modified in MPLCs.} As the results of RQ7 revealed, the source files modified in MPLCs are likely to have a higher defect density than that of source files only modified in non-MPLCs. Therefore, practitioners should pay more attention to the former. For instance, designers may improve the modularity of source files modified in MPLCs, in order to lower the likelihood of such source files being modified together; developers need to investigate deeper on the impact of MPLCs; and developers and testers can invest more effort to test such source files.

\subsection{Implications for Researchers}
\textbf{Take MPLCs into account when constructing defect prediction models for MPL software systems.} Since whether the source files have been modified in MPLCs plays a role in the defect density of source files, it is reasonable to take MPLCs as a factor in defect prediction for MPL systems.

\textbf{Investigate the factors that influence the proportion of MPLCs in MPL software systems.} The change complexity of MPLCs is much higher than that of non-MPLCs, which implies that MPLCs will greatly increase the development cost of MPL software systems. Therefore, it is necessary to keep the proportion of MPLCs of an MPL system under a reasonable level. However, it still remains unclear what factors contribute to a relatively high proportion of MPLCs in an MPL system, which is an interesting research question.

\textbf{Require larger-scale studies.} As shown in Table \ref{table:DemographicInformation}, some PLs (e.g., Clojure, Kotlin, and Swift) are seldom used in the selected projects, while some other PLs (e.g., Java, Python, C/C++, and JavaScript) are commonly used in the selected projects. More projects written by the seldom used PLs are needed to balance the use of different PLs. To our experience, it is not possible to find many more Apache projects for research on MPLCs. Instead, it is better to turn to a bigger platform, such as GitHub, to find more suitable projects. 

\textbf{Need further studies on the relationship between MPLCs and software architecture.} Although intuitively MPLCs are more related to the changes at the architecture level, there still lacks evidence on how MPLCs relate to software architecture and vice verse. We believe that this could be a promising research topic to be further explored. 

\textbf{Explore the reasons for why a majority of MPLCs introduce bugs in specific projects.} For instance, the project \textit{Ambari}, the percentage of bug-introducing MPLCs is more than 85\%, an incredibly high level. In such a software system, it is highly risky to make every change involving source files written in multiple PLs. Thus, it is valuable to investigate the reasons behind this phenomenon, so as to take actions to avoid a large percentage of bug-introducing MPLCs.

\section{Threats to Validity}\label{chap:threats}

There are several threats to the validity of the study results. We discuss these threats according to the guidelines in \citep{RuHo2009}. Please note that internal validity is not discussed, since we do not study causal relationships.

\subsection{Construct Validity}
Construct validity is concerned with whether the values of the variables (listed in Table \ref{table:DataItemsForCommit} and Table \ref{table:DataItemsForFile}) we obtained are in line with the real values that we expected. A potential threat to construct validity is that not all issues resolved are linked to corresponding commits. Due to different developer habits and development cultures, committers may not explicitly mention the ID of the issue resolved in corresponding commit message, which may negatively affect the representativeness of the collected issues and further influence the accuracy of defect density and the time taken to resolve issues. Through our analysis (the analysis results are not shown in this paper due to its deviation from the focus of this paper), we confirmed that the committers who explicitly mention the issue ID do not come from a small group of specific developers. Therefore, this threat is to some extent mitigated. 

Another potential threat is related to the commits corresponding to the MERGE operation. Since a MERGE commit is usually a combination of a number of commits from different branches of the repository, such a commit is likely to be an MPLC in an MPL software system. Furthermore, the changed source files in the MERGE commit are duplicate with those in the merged commits. If such MERGE commits are still kept in the dataset, the quality of the dataset would be negatively affected, especially when a project has frequent MERGE operations. As described in Section \ref{datacollectionprocedure}, we discarded the MERGE commits, and thus this threat is totally removed. 

One more threat is that a bug's bug-introducing commits identified by the SZZ method implemented in the PyDriller tool may not be the real cause of the bug. Abidi et al. manually evaluated the effectiveness of the SZZ method implemented in PyDriller, and found that the precision of bug-introducing commits identified by PyDriller is more than 70\% \citep{AbRaOpKh2021}. Thus, this threat is partially mitigated.

Finally, the open time of an issue is considered as the time between its creation and resolution without removing the ``assignment time'' between its creation and it being assigned to a developer, which may cause the inaccuracy of open time calculation. In our study, we did not remove the assignment time, since (1) our dataset (manually exported from JIRA) does not contain the time information when an issue is assigned to a developer, (2) some closed or resolved issues even do not have any developers assigned to them, and thus there is no assignment time for such issues, and (3) it is possible that an issue is not assigned to a developer until it is actually resolved by the developer, and removing the assignment time of such issues will reduce their actual open time.

\subsection{External Validity}
External validity is concerned with the generalizability of the study results. First, a potential threat to external validity is whether the selected projects are representative enough. As presented in Section \ref{CaseSelection}, we applied a set of criteria to select projects. We tried to include as many as possible the Apache projects that meet the selection criteria. Furthermore, the selected projects cover different application domains, and differ in code repository size and development duration. This indicates improved representativeness of the selected projects.

Second, another threat is that only Apache MPL OSS projects were selected. The number of available projects is relatively small, which may reduce the generalizability of the study results. 

Third, the use of different PLs by the selected projects are not balanced, which exposes a threat to the external validity. Some PLs are used by all selected project, while some other PLs are used by only a couple of selected projects. This threat can be mitigated by including more projects using the PLs that are not adopted in the selected projects in this paper. This is a direction of our future work.

Finally, since only 18 PLs are considered in this study, the findings and the conclusions drawn are only valid for projects using these PLs. Since only OSS projects were selected, we cannot generalize the findings and conclusions to closed source software projects.

\subsection{Reliability}
Reliability refers to whether the study yields the same results when it is replicated by other researchers. A potential threat is related to the implementation of related software tools for data collection. The tools were mainly implemented by the third author, and the code of the key functionalities had been regularly reviewed by the first and second authors. In addition, sufficient tests were performed to ensure the correctness of the calculation of data items. Hence, the threat to reliability had been alleviated.

Another threat is related to the correctness of the Mann-Whitney U tests. Since we only used IBM SPSS (a well-engineered and widely-used professional tool for statistics) to run the tests, this threat is minimal.

\section{Conclusions and Future Work}
\label{conclusions}

The phenomenon of MPLCs is prevalent in modern software system development. To our knowledge, this phenomenon has not been explored yet. In light of the potential influence of MPLCs on development difficulty and software quality, we conducted an empirical study to understand the state of MPLCs, their change complexity, as well as their impacts on open time of issues and bug proneness of source files in real-life software projects.  

Following a set of predefined criteria, we selected 18 non-trivial Apache MPL OSS projects, in which 197,566 commits (including 18,372 MPLCs) were analyzed. The main findings are that:

\begin{itemize}
 \setlength{\itemsep}{0pt}
 \setlength{\parsep}{0pt}
  \item The most commonly used PL combination in the selected projects consists of all of the four PLs, i.e., C/C++, Java, JavaScript, and Python.
  \item The proportion of MPLCs for the selected projects ranges from 1.7\% to 41.0\%, and 9.3\% of the commits are MPLCs when taking all projects as a whole. The proportion of MPLCs goes to a relatively stable level for 83\% of the selected projects.
  \item In most of the selected projects, the average number of PLs used in each MPLC is around 2.0. Particularly, when taking all selected projects as a whole, 91.7\% of the MPLCs involve source files written in two programming languages. 
  \item The change complexity (in terms of the number of lines of code, source files, directories modified, and Entropy) of MPLCs is significantly higher than that of non-MPLCs in all selected projects. 
  \item In 89\% of the selected projects, the issues fixed in MPLCs take longer (by 8.0\% to 124.7\%) to be resolved than the issues fixed in non-MPLCs.
  \item In 89\% of the selected projects, there are no significant differences between MPLCs and non-MPLCs regarding the impact on issue reopen. 
  \item In 83\% of the selected projects, the source files that have been modified in MPLCs are more bug-prone (by 17.2\% to 3,681.8\%) in terms of defect density than source files that have never been modified in MPLCs.
  \item MPLCs are more likely to introduce bugs than non-MPLCs.
\end{itemize}

Based on the results of this exploratory study, our future research will focus on the following directions: First, one promising direction is to investigate how MPLCs and software architecture interplay in MPL software systems. For instance, we may look into whether MPLCs are related to architectural technical debt \citep{LiLiAv2015,liAvLi2015}.
Second, we plan to use MPLCs as an additional factor to enhance bug prediction models for MPL systems based on existing models.
Third, to better understand how bugs are introduced in MPLCs and non-MPLCs, we will conduct an in-depth analysis on several selected projects.
Finally, we will replicate this study by constructing a large-scale dataset, in which more projects written in diverse PLs will be included to balance the use of different PLs.

\section*{Acknowledgment}
This work is supported by the Natural Science Foundation of Hubei Province of China under Grant No. 2021CFB577, and the National Natural Science Foundation of China under Grant Nos. 62176099, 62172311, and 61702377.

\printcredits

\bibliographystyle{cas-model2-names}

\bibliography{references}

\begin{thebibliography}{32}
\expandafter\ifx\csname natexlab\endcsname\relax\def\natexlab#1{#1}\fi
\providecommand{\url}[1]{\texttt{#1}}
\providecommand{\href}[2]{#2}
\providecommand{\path}[1]{#1}
\providecommand{\DOIprefix}{doi:}
\providecommand{\ArXivprefix}{arXiv:}
\providecommand{\URLprefix}{URL: }
\providecommand{\Pubmedprefix}{pmid:}
\providecommand{\doi}[1]{\href{http://dx.doi.org/#1}{\path{#1}}}
\providecommand{\Pubmed}[1]{\href{pmid:#1}{\path{#1}}}
\providecommand{\bibinfo}[2]{#2}
\ifx\xfnm\relax \def\xfnm[#1]{\unskip,\space#1}\fi
%Type = Inproceedings
\bibitem[{Abidi et~al.(2019a)Abidi, Grichi and Khomh}]{AbGrKh2019}
\bibinfo{author}{Abidi, M.}, \bibinfo{author}{Grichi, M.},
  \bibinfo{author}{Khomh, F.}, \bibinfo{year}{2019}a.
\newblock \bibinfo{title}{Behind the scenes: developers' perception of
  multi-language practices}, in: \bibinfo{booktitle}{Proceedings of the 29th
  Annual International Conference on Computer Science and Software Engineering
  (CASCON'19)}, \bibinfo{publisher}{IBM Corp.}. pp. \bibinfo{pages}{72--81}.
%Type = Inproceedings
\bibitem[{Abidi et~al.(2019b)Abidi, Grichi, Khomh and
  Guéhéneuc}]{AbKhGu2019b}
\bibinfo{author}{Abidi, M.}, \bibinfo{author}{Grichi, M.},
  \bibinfo{author}{Khomh, F.}, \bibinfo{author}{Guéhéneuc, Y.G.},
  \bibinfo{year}{2019}b.
\newblock \bibinfo{title}{Code smells for multi-language systems}, in:
  \bibinfo{booktitle}{Proceedings of the 24th European Conference on Pattern
  Languages of Programs (EuroPLoP'19)}, \bibinfo{publisher}{ACM}. p.
  \bibinfo{pages}{Article 12}.
%Type = Inproceedings
\bibitem[{Abidi et~al.(2019c)Abidi, Khomh and Guéhéneuc}]{AbKhGu2019a}
\bibinfo{author}{Abidi, M.}, \bibinfo{author}{Khomh, F.},
  \bibinfo{author}{Guéhéneuc, Y.G.}, \bibinfo{year}{2019}c.
\newblock \bibinfo{title}{Anti-patterns for multi-language systems}, in:
  \bibinfo{booktitle}{Proceedings of the 24th European Conference on Pattern
  Languages of Programs (EuroPLoP'19)}, \bibinfo{publisher}{ACM}. p.
  \bibinfo{pages}{Article 42}.
%Type = Article
\bibitem[{Abidi et~al.(2021)Abidi, Rahman, Openja and Khomh}]{AbRaOpKh2021}
\bibinfo{author}{Abidi, M.}, \bibinfo{author}{Rahman, M.S.},
  \bibinfo{author}{Openja, M.}, \bibinfo{author}{Khomh, F.},
  \bibinfo{year}{2021}.
\newblock \bibinfo{title}{Are multi-language design smells fault-prone? an
  empirical study}.
\newblock \bibinfo{journal}{ACM Transactions on Software Engineering and
  Methodology} \bibinfo{volume}{30}, \bibinfo{pages}{Article No. 29}.
%Type = Misc
\bibitem[{Basili(1992)}]{Ba1992}
\bibinfo{author}{Basili, V.R.}, \bibinfo{year}{1992}.
\newblock \bibinfo{title}{Software modeling and measurement: The
  goal/question/metric paradigm}.
\newblock \URLprefix
  \url{http://drum.lib.umd.edu/bitstream/1903/7538/1/Goal_Question_Metric.pdf}.
%Type = Article
\bibitem[{Benjamini and Hochberg(1995)}]{BeHo1995}
\bibinfo{author}{Benjamini, Y.}, \bibinfo{author}{Hochberg, Y.},
  \bibinfo{year}{1995}.
\newblock \bibinfo{title}{Controlling the false discovery rate: A practical and
  powerful approach to multiple testing}.
\newblock \bibinfo{journal}{Journal of the Royal Statistical Society: Series B
  (Methodological)} \bibinfo{volume}{57}, \bibinfo{pages}{289--300}.
%Type = Article
\bibitem[{Berger et~al.(2019)Berger, Hollenbeck, Maj, Vitek and
  Vitek}]{BeHoMaViVi2019}
\bibinfo{author}{Berger, E.D.}, \bibinfo{author}{Hollenbeck, C.},
  \bibinfo{author}{Maj, P.}, \bibinfo{author}{Vitek, O.},
  \bibinfo{author}{Vitek, J.}, \bibinfo{year}{2019}.
\newblock \bibinfo{title}{On the impact of programming languages on code
  quality: A reproduction study}.
\newblock \bibinfo{journal}{ACM Transactions on Programming Languages and
  Systems} \bibinfo{volume}{41}, \bibinfo{pages}{Article 21}.
%Type = Inproceedings
\bibitem[{Bhattacharya and Neamtiu(2011)}]{BhNe2011}
\bibinfo{author}{Bhattacharya, P.}, \bibinfo{author}{Neamtiu, I.},
  \bibinfo{year}{2011}.
\newblock \bibinfo{title}{Assessing programming language impact on development
  and maintenance: a study on {C} and {C}++}, in:
  \bibinfo{booktitle}{Proceedings of the 33rd International Conference on
  Software Engineering (ICSE'11)}, pp. \bibinfo{pages}{171--180}.
%Type = Book
\bibitem[{Field(2013)}]{Fi2013}
\bibinfo{author}{Field, A.}, \bibinfo{year}{2013}.
\newblock \bibinfo{title}{Discovering Statistics using IBM SPSS Statistics}.
\newblock \bibinfo{edition}{Fourth} ed., \bibinfo{publisher}{Sage Publications
  Ltd.}, \bibinfo{address}{Singapore}.
%Type = Article
\bibitem[{Grichi et~al.(2021)Grichi, Abidi, Jaafar, Eghan and
  Adams}]{GrAbJa2020}
\bibinfo{author}{Grichi, M.}, \bibinfo{author}{Abidi, M.},
  \bibinfo{author}{Jaafar, F.}, \bibinfo{author}{Eghan, E.E.},
  \bibinfo{author}{Adams, B.}, \bibinfo{year}{2021}.
\newblock \bibinfo{title}{On the impact of interlanguage dependencies in
  multilanguage systems empirical case study on java native interface
  applications {(JNI)}}.
\newblock \bibinfo{journal}{IEEE Transactions on Reliability}
  \bibinfo{volume}{70}, \bibinfo{pages}{428--440}.
%Type = Inproceedings
\bibitem[{Grichi et~al.(2020)Grichi, Eghan and Adams}]{GrEgAd2020}
\bibinfo{author}{Grichi, M.}, \bibinfo{author}{Eghan, E.E.},
  \bibinfo{author}{Adams, B.}, \bibinfo{year}{2020}.
\newblock \bibinfo{title}{On the impact of multi-language development in
  machine learning frameworks}, in: \bibinfo{booktitle}{Proceedings of the 36th
  IEEE International Conference on Software Maintenance and Evolution
  (ICSME'20)}, \bibinfo{publisher}{IEEE}. pp. \bibinfo{pages}{546--556}.
%Type = Inproceedings
\bibitem[{Hassan(2009)}]{Ha2009}
\bibinfo{author}{Hassan, A.E.}, \bibinfo{year}{2009}.
\newblock \bibinfo{title}{Predicting faults using the complexity of code
  changes}, in: \bibinfo{booktitle}{Proceedings of the 31st International
  Conference on Software Engineering (ICSE'09)}, \bibinfo{publisher}{IEEE}. pp.
  \bibinfo{pages}{78--88}.
%Type = Article
\bibitem[{Kargar et~al.(2019)Kargar, Isazadeh and Izadkhah}]{kaIsIz2019}
\bibinfo{author}{Kargar, M.}, \bibinfo{author}{Isazadeh, A.},
  \bibinfo{author}{Izadkhah, H.}, \bibinfo{year}{2019}.
\newblock \bibinfo{title}{Multi-programming language software systems
  modularization}.
\newblock \bibinfo{journal}{Computers \& Electrical Engineering}
  \bibinfo{volume}{80}, \bibinfo{pages}{106500}.
%Type = Article
\bibitem[{Kargar et~al.(2020)Kargar, Isazadeh and Izadkhah}]{kaIsIz2020}
\bibinfo{author}{Kargar, M.}, \bibinfo{author}{Isazadeh, A.},
  \bibinfo{author}{Izadkhah, H.}, \bibinfo{year}{2020}.
\newblock \bibinfo{title}{Improving the modularization quality of heterogeneous
  multi-programming software systems by unifying structural and semantic
  concepts}.
\newblock \bibinfo{journal}{Journal of Supercomputing} \bibinfo{volume}{76},
  \bibinfo{pages}{87--121}.
%Type = Inproceedings
\bibitem[{Kochhar et~al.(2016)Kochhar, Wijedasa and Lo}]{KoWiLo2016}
\bibinfo{author}{Kochhar, P.S.}, \bibinfo{author}{Wijedasa, D.},
  \bibinfo{author}{Lo, D.}, \bibinfo{year}{2016}.
\newblock \bibinfo{title}{A large scale study of multiple programming languages
  and code quality}, in: \bibinfo{booktitle}{Proceedings of the IEEE 23rd
  International Conference on Software Analysis, Evolution, and Reengineering
  (SANER'16)}, pp. \bibinfo{pages}{563--573}.
%Type = Inproceedings
\bibitem[{Kontogiannis et~al.(2006)Kontogiannis, Linos and Wong}]{KoLiWo2006}
\bibinfo{author}{Kontogiannis, K.}, \bibinfo{author}{Linos, P.},
  \bibinfo{author}{Wong, K.}, \bibinfo{year}{2006}.
\newblock \bibinfo{title}{Comprehension and maintenance of large-scale
  multi-language software applications}, in: \bibinfo{booktitle}{Proceedings of
  the 22nd IEEE International Conference on Software Maintenance (ICSM'06)},
  \bibinfo{publisher}{IEEE}. pp. \bibinfo{pages}{497--500}.
%Type = Article
\bibitem[{Li et~al.(2015a)Li, Avgeriou and Liang}]{liAvLi2015}
\bibinfo{author}{Li, Z.}, \bibinfo{author}{Avgeriou, P.},
  \bibinfo{author}{Liang, P.}, \bibinfo{year}{2015}a.
\newblock \bibinfo{title}{A systematic mapping study on technical debt and its
  management}.
\newblock \bibinfo{journal}{Journal of Systems and Software}
  \bibinfo{volume}{101}, \bibinfo{pages}{193--220}.
%Type = Inproceedings
\bibitem[{Li et~al.(2015b)Li, Liang and Avgeriou}]{LiLiAv2015}
\bibinfo{author}{Li, Z.}, \bibinfo{author}{Liang, P.},
  \bibinfo{author}{Avgeriou, P.}, \bibinfo{year}{2015}b.
\newblock \bibinfo{title}{Architectural technical debt identification based on
  architecture decisions and change scenarios}, in:
  \bibinfo{booktitle}{Proceedings of the 12th Working IEEE/IFIP Conference on
  Software Architecture (WICSA'15)}, \bibinfo{publisher}{IEEE}. pp.
  \bibinfo{pages}{65--74}.
%Type = Inproceedings
\bibitem[{Li et~al.(2017)Li, Liang and Li}]{LiLiLi2017}
\bibinfo{author}{Li, Z.}, \bibinfo{author}{Liang, P.}, \bibinfo{author}{Li,
  B.}, \bibinfo{year}{2017}.
\newblock \bibinfo{title}{Relating alternate modifications to defect density in
  software development}, in: \bibinfo{booktitle}{Proceedings of the 39th
  International Conference on Software Engineering Companion (ICSE'17)},
  \bibinfo{publisher}{IEEE}. pp. \bibinfo{pages}{308--310}.
%Type = Article
\bibitem[{Li et~al.(2020)Li, Liang, Li, Mo and Li}]{LiLiLiMoLi2020}
\bibinfo{author}{Li, Z.}, \bibinfo{author}{Liang, P.}, \bibinfo{author}{Li,
  D.}, \bibinfo{author}{Mo, R.}, \bibinfo{author}{Li, B.},
  \bibinfo{year}{2020}.
\newblock \bibinfo{title}{Is bug severity in line with bug fixing change
  complexity?}
\newblock \bibinfo{journal}{International Journal of Software Engineering and
  Knowledge Engineering} \bibinfo{volume}{30}, \bibinfo{pages}{1779--1800}.
%Type = Inproceedings
\bibitem[{Li et~al.(2021)Li, Qi, Yu, Liang, Mo and Yang}]{LiQiYu2021}
\bibinfo{author}{Li, Z.}, \bibinfo{author}{Qi, X.}, \bibinfo{author}{Yu, Q.},
  \bibinfo{author}{Liang, P.}, \bibinfo{author}{Mo, R.}, \bibinfo{author}{Yang,
  C.}, \bibinfo{year}{2021}.
\newblock \bibinfo{title}{Multi-programming-language commits in oss: An
  empirical study on apache projects}, in: \bibinfo{booktitle}{Proceedings of
  the 29th IEEE/ACM International Conference on Program Comprehension
  (ICPC'21)}, \bibinfo{publisher}{IEEE}. pp. \bibinfo{pages}{1--11}.
%Type = Article
\bibitem[{Li et~al.(2022)Li, Qi, Yu, Liang, Mo and Yang}]{studymaterial}
\bibinfo{author}{Li, Z.}, \bibinfo{author}{Qi, X.}, \bibinfo{author}{Yu, Q.},
  \bibinfo{author}{Liang, P.}, \bibinfo{author}{Mo, R.}, \bibinfo{author}{Yang,
  C.}, \bibinfo{year}{2022}.
\newblock \bibinfo{title}{Dataset for "{Exploring Multi-Programming-Language
  Commits and Their Impacts on Software Quality: An Empirical Study on Apache
  Projects}"}.
\newblock \bibinfo{journal}{https://github.com/ASSMS/JSS/tree/main/Dataset} .
%Type = Article
\bibitem[{Mayer(2017)}]{Ma2017}
\bibinfo{author}{Mayer, P.}, \bibinfo{year}{2017}.
\newblock \bibinfo{title}{A taxonomy of cross-language linking mechanisms in
  open source frameworks}.
\newblock \bibinfo{journal}{Computing} \bibinfo{volume}{99},
  \bibinfo{pages}{701--724}.
%Type = Inproceedings
\bibitem[{Mayer and Bauer(2015)}]{MaBa2015}
\bibinfo{author}{Mayer, P.}, \bibinfo{author}{Bauer, A.}, \bibinfo{year}{2015}.
\newblock \bibinfo{title}{An empirical analysis of the utilization of multiple
  programming languages in open source projects}, in:
  \bibinfo{booktitle}{Proceedings of the 19th International Conference on
  Evaluation and Assessment in Software Engineering (EASE'15)},
  \bibinfo{publisher}{ACM}. p. \bibinfo{pages}{Article 4}.
%Type = Article
\bibitem[{Mayer et~al.(2017)Mayer, Kirsch and Le}]{MaKiLe2017}
\bibinfo{author}{Mayer, P.}, \bibinfo{author}{Kirsch, M.}, \bibinfo{author}{Le,
  M.A.}, \bibinfo{year}{2017}.
\newblock \bibinfo{title}{On multi-language software development,
  cross-language links and accompanying tools: a survey of professional
  software developers}.
\newblock \bibinfo{journal}{Journal of Software Engineering Research and
  Development} \bibinfo{volume}{5}, \bibinfo{pages}{1--33}.
%Type = Inproceedings
\bibitem[{Mo et~al.(2016)Mo, Cai, Kazman, Xiao and Feng}]{MoCaKaXiFe2016}
\bibinfo{author}{Mo, R.}, \bibinfo{author}{Cai, Y.}, \bibinfo{author}{Kazman,
  R.}, \bibinfo{author}{Xiao, L.}, \bibinfo{author}{Feng, Q.},
  \bibinfo{year}{2016}.
\newblock \bibinfo{title}{Decoupling level: A new metric for architectural
  maintenance complexity}, in: \bibinfo{booktitle}{Proceedings of the 38th
  International Conference on Software Engineering (ICSE'16)},
  \bibinfo{publisher}{ACM}. pp. \bibinfo{pages}{499--510}.
%Type = Inproceedings
\bibitem[{Ray et~al.(2014)Ray, Posnett, Filkov and Devanbu}]{RaPoFiDe2014}
\bibinfo{author}{Ray, B.}, \bibinfo{author}{Posnett, D.},
  \bibinfo{author}{Filkov, V.}, \bibinfo{author}{Devanbu, P.},
  \bibinfo{year}{2014}.
\newblock \bibinfo{title}{A large scale study of programming languages and code
  quality in github}, in: \bibinfo{booktitle}{Proceedings of the 22nd ACM
  SIGSOFT International Symposium on Foundations of Software Engineering
  (FSE'14)}, \bibinfo{publisher}{ACM}. pp. \bibinfo{pages}{155--165}.
%Type = Article
\bibitem[{Runeson and H{\"o}st(2009)}]{RuHo2009}
\bibinfo{author}{Runeson, P.}, \bibinfo{author}{H{\"o}st, M.},
  \bibinfo{year}{2009}.
\newblock \bibinfo{title}{Guidelines for conducting and reporting case study
  research in software engineering}.
\newblock \bibinfo{journal}{Empirical Software Engineering}
  \bibinfo{volume}{14}, \bibinfo{pages}{131--164}.
%Type = Misc
\bibitem[{Shatnawi et~al.(2019)Shatnawi, Mili, Abdellatif, Guéhéneuc, Moha,
  Hecht, El~Boussaidi and Privat}]{ShMiAb2019}
\bibinfo{author}{Shatnawi, A.}, \bibinfo{author}{Mili, H.},
  \bibinfo{author}{Abdellatif, M.}, \bibinfo{author}{Guéhéneuc, Y.G.},
  \bibinfo{author}{Moha, N.}, \bibinfo{author}{Hecht, G.},
  \bibinfo{author}{El~Boussaidi, G.}, \bibinfo{author}{Privat, J.},
  \bibinfo{year}{2019}.
\newblock \bibinfo{title}{Static code analysis of multilanguage software
  systems}.
\newblock \URLprefix \url{https://arxiv.org/abs/arXiv:1906.00815v1}.
%Type = Inproceedings
\bibitem[{Spadini et~al.(2018)Spadini, Aniche and Bacchelli}]{Spadini2018}
\bibinfo{author}{Spadini, D.}, \bibinfo{author}{Aniche, M.},
  \bibinfo{author}{Bacchelli, A.}, \bibinfo{year}{2018}.
\newblock \bibinfo{title}{{PyDriller: Python framework for mining software
  repositories}}, in: \bibinfo{booktitle}{Proceedings of the 26th ACM Joint
  Meeting on European Software Engineering Conference and Symposium on the
  Foundations of Software Engineering (ESEC/FSE'18)}, \bibinfo{publisher}{ACM}.
  pp. \bibinfo{pages}{908--911}.
%Type = Inproceedings
\bibitem[{Wong et~al.(2011)Wong, Cai, Kim and Dalton}]{WoCaKiDa2011}
\bibinfo{author}{Wong, S.}, \bibinfo{author}{Cai, Y.}, \bibinfo{author}{Kim,
  M.}, \bibinfo{author}{Dalton, M.}, \bibinfo{year}{2011}.
\newblock \bibinfo{title}{Detecting software modularity violations}, in:
  \bibinfo{booktitle}{Proceedings of the 33rd International Conference on
  Software Engineering (ICSE'11)}, \bibinfo{publisher}{IEEE}. pp.
  \bibinfo{pages}{411--420}.
%Type = Article
\bibitem[{Śliwerski et~al.(2005)Śliwerski, Zimmermann and Zeller}]{SZZ2005}
\bibinfo{author}{Śliwerski, J.}, \bibinfo{author}{Zimmermann, T.},
  \bibinfo{author}{Zeller, A.}, \bibinfo{year}{2005}.
\newblock \bibinfo{title}{When do changes induce fixes?}
\newblock \bibinfo{journal}{ACM SIGSOFT Software Engineering Notes}
  \bibinfo{volume}{30}, \bibinfo{pages}{1--5}.

\end{thebibliography}

\balance

\end{sloppypar}
\end{document}